\documentclass[10pt,twoside]{article}

\usepackage{working-paper}
\usepackage{booktabs,tabularx}
\usepackage{graphicx,placeins}
\usepackage{natbib}
\usepackage{xurl}
\usepackage[hidelinks]{hyperref}

\DeclareMathOperator{\clip}{clip}

\newtheoremstyle{paperplain}{8pt}{8pt}{\itshape}{}{\bfseries}{.}{0.5em}
  {\thmname{#1}\thmnumber{ #2}\thmnote{\normalfont\space(#3)}}
\newtheoremstyle{paperdefinition}{8pt}{8pt}{\normalfont}{}{\itshape}{.}{0.5em}
  {\thmname{#1}\thmnumber{ #2}\thmnote{\normalfont\space(#3)}}
\theoremstyle{paperplain}
\newtheorem{theorem}{Theorem}
\newtheorem{proposition}[theorem]{Proposition}
\newtheorem{lemma}[theorem]{Lemma}
\newtheorem{corollary}[theorem]{Corollary}
\theoremstyle{paperdefinition}

\newtheorem{remark}{Remark}

\allowdisplaybreaks[1]

\title{Strategic Repayment on Graphons:\\
Which Payment Freezes Survive Network Sampling}
\author{Hicham El Kaissi\thanks{Hicham El Kaissi: University of Kent
  (email: \href{mailto:h.el-kaissi@kent.ac.uk}{h.el-kaissi@kent.ac.uk}).}}
\date{November 2026}
\hypersetup{pdftitle={Strategic Repayment on Graphons: Which Payment Freezes Survive Network Sampling},pdfauthor={Hicham El Kaissi}}

\begin{document}
\maketitle

\begin{abstract}
In the two-date settlement model of \citet{elkaissi2026}, banks can defer
payments they can afford to retain cash for an interim funding need. We
extend the game to directed graphons to assess settlement risk when
bilateral liabilities are unobserved. In a four-bank economy, the least
strategic equilibrium repays between \(50\%\) and \(100\%\) across networks
with identical balance-sheet totals. Both bounds are attained, and
mechanical clearing uniquely gives full payment throughout. For bounded
kernels and growing Dirichlet concentration, equilibria of Sparsemax--Dirichlet
networks approach the graphon equilibrium set. Strict repayment margins and
an invertible interior equilibrium equation give exact finite equilibria
near individual graphon equilibria, including some noncontractive branches.
A finite-sample probability bound gives conditions for a freeze and full
early payment to coexist. In an acyclic economy with full terminal settlement,
fixed concentration creates a persistent early-payment shortfall, which we
derive exactly. At unit concentration it is approximately \(14.04\)
percentage points.
\par\vspace{6pt}
{\small (JEL C72, D85, E42, G21)\par}
\vspace{2pt}
{\normalfont\footnotesize\textit{Keywords:} directed graphons, financial networks,
strategic repayment, payment freezes, network reconstruction, equilibrium multiplicity.\par}
\end{abstract}

\section{Introduction}
A bank can have enough cash to pay and still choose to wait. In the
two-date settlement model of \citet{elkaissi2026}, banks weigh the cost of
deferring principal against the value of retaining cash for an uncertain
funding need. The need arises after early settlement and before deferred
principal falls due. When a counterparty misses a payment, the creditor
loses cash and may want to hold more in reserve. If both effects operate
and repayment stays interior, a pound of missed receipts cuts the
creditor's own early payments by more than a pound. Deferred principal
can still be paid in full at the final date.

How much can an analyst infer about early repayment from bank-level
liabilities and claims when bilateral exposures are unobserved? In El Kaissi's
finite settlement model, those totals, opening liquidity and reserve
parameters give tests for system-wide zero and full early payment.
Proposition~\ref{samp:fixed-margin-corners} establishes that the tests are
invariant across feasible reconstructions. At a mixed repayment profile,
receipts depend on which debtors pay. In the four-bank example of
Section~\ref{ex:matched-marginals}, the least strategic repayment rate
has sharp bounds of \(50\%\) and \(100\%\) across feasible reconstructions.
Every reconstruction has unique full mechanical clearing.

Feasible reconstructions change bilateral exposures while matching both
balance-sheet margins. Sparsemax--Dirichlet sampling varies allocations
around a fitted directed kernel, preserving each debtor's total obligations
and letting creditor claims vary. Sparsemax selects and reweights creditor
links. Dirichlet draws vary the shares on that support. The step-kernel
embedding preserves early cash flows, reserve choices and equilibria exactly.

For bounded kernels and growing Dirichlet concentration, every finite
equilibrium approaches the equilibrium set of the target graphon. Strict
repayment margins and invertibility of the interior equilibrium equation
give an exact finite counterpart to a graphon equilibrium, even for some
branches with feedback too strong for contraction. The equilibrium sets
converge in Hausdorff distance when every graphon equilibrium satisfies
these conditions. An explicit finite-sample
condition certifies coexistence of a freeze and full early payment,
together with unique full mechanical clearing. In an acyclic economy with a unique graphon
equilibrium, fixed concentration leaves dispersion in sampled receipts and
the payment cap turns it into a persistent shortfall. Theorem~\ref{ex:fixed-limit}
derives the limiting shortfall, which is approximately \(14.04\) percentage
points at unit concentration.

\paragraph{Related literature.}\label{sec:related}

Proportional clearing links payments to realised resources
\citep{eisenbergnoe2001}. Dependency cycles can sustain self-fulfilling
insolvency when default is costly \citep{jacksonpernoud2024}. Precautionary
reserve choice allows an early payment freeze to coexist with unique full
mechanical clearing.

Graph limits and exchangeable sampling motivate the graphon representation \citep{lovaszszegedy2006,diaconisjanson2008}. Graphon equilibria approximate sampled-network equilibria under uniqueness
\citep{pariseozdaglar2023}. With multiplicity, existing results identify limits
of convergent finite equilibria and approximate graphon equilibria by finite
\(\varepsilon\)-equilibria \citep{rokadeparise2026}. Strict repayment margins
and an invertible interior equilibrium equation yield exact nearby finite
equilibria in our game. Related work uses graphons to study threshold contagion and
extremal solvency cutoffs
\citep{erolpariseteytelboym2023,dasarathavenkateshvohra2026}.
\citet{feng2026} establishes sampled-network convergence for dynamic
distress paths under a condition on the mass near the distress threshold.

\section{The strategic settlement model}
\label{mod:game}

The directed graphon embedding of the two-date settlement game of
\citet{elkaissi2026} preserves early cash flows, reserve choices and
equilibria for any finite liability network.

\subsection{Early repayment and terminal settlement}
\label{mod:criterion}

At date 0, banks simultaneously choose how much of their interbank
principal to repay out of opening liquidity and incoming payments.
They hold the remaining cash as a reserve against an uncertain funding
need that arises before deferred principal falls due. At date 1, cash
released from temporary interim use and other retained reserves, together
with continuation cash, are available to settle deferred principal.

Holding cash back can save a bank the cost of covering an interim
shortfall, but another pound kept in reserve means another pound of
principal deferred. A missed receipt leaves the creditor with less cash
and can increase the scale of its interim funding need. A creditor that
responds by paying less passes this pressure on to its own creditors.

\paragraph{Cash, reserves and funding needs.}
In an \(n\)-bank network, \(L_{ij}\geq0\) is bank \(i\)'s nominal liability to
bank \(j\), with \(L_{ii}=0\), \(\bar p_i=\sum_jL_{ij}\) and
\(\bar z_i=\sum_jL_{ji}\). Let \(A_n=\{i:\bar p_i>0\}\) be the active debtors.
For \(i\in A_n\), set \(\Pi_{ij}=L_{ij}/\bar p_i\). The other rows are zero.
A repayment profile \(s\in[0,1]^n\) specifies early repayment fractions, with
\(s_j=0\) for \(j\notin A_n\). Bank \(i\) receives \(z_i(s)=\sum_j L_{ji}s_j\).

An active bank \(i\) takes its counterparties' payments as given when
choosing its own early repayment fraction \(u\). Its net unrestricted external
liquidity \(e_i\) accounts for senior near-term obligations outside the
network and can be negative. For conjecture \(s\), define missed receipts
\(\Delta_i\), available cash \(a_i\), funding need scale \(\kappa_i\), and the
cash reserve \(R_i\) left after payment.
\begin{align}
 \Delta_i(s)&=\bar z_i-z_i(s),&
 a_i(s)&=[e_i+z_i(s)]_+,\notag\\
 \kappa_i(s)&=\omega_i+\psi_i\Delta_i(s),&
 R_i(u\mid s)&=a_i(s)-\bar p_i u .
 \label{mod:resources}
\end{align}
Here \([r]_+=\max\{0,r\}\), and feasibility requires
\(0\leq u\leq\min\{1,a_i(s)/\bar p_i\}\).
The funding need is \(\kappa_i(s)Y_i\), with \(\omega_i>0\), \(\psi_i\geq0\)
and a nonnegative, profile-invariant shock \(Y_i\) of finite mean.
Its survival function \(\overline G_i(t)=\Pr(Y_i>t)\) is continuous and
strictly decreasing, with \(\overline G_i(0)=1\) and limit zero as \(t\to\infty\).

\paragraph{The date-0 objective.}
The bank faces two costs. It pays \(\chi_i>0\) per unit of interim funding
need that reserves leave uncovered. It also bears a cost
\(\delta_i=d_i^{\mathrm{priv}}+k_i>0\) per unit of deferred principal,
comprising a private continuation-value
cost \(d_i^{\mathrm{priv}}\) and a late-interest charge \(k_i\geq0\).
The bank maximises
\begin{equation}
 U_i^{\mathrm{net}}(u\mid s)=
 -\chi_i\mathbb E\!\left[\big(\kappa_i(s)Y_i-R_i(u\mid s)\big)_+\right]
 -\delta_i\bar p_i(1-u).
 \label{mod:netpayoff}
\end{equation}
Interim funding expenses enter the payoff outside the final settlement
account. The bank pledges at least \(k_i\bar p_i\) in cash to cover interest
even if it defers all principal. This pledge is excluded from unrestricted
liquidity at both dates.

\paragraph{Optimal reserves and early repayment.}
An extra pound in reserve saves \(\chi_i\) only when the funding need
exceeds reserves. The desired reserve sets this expected marginal saving
equal to the cost \(\delta_i\) when \(\delta_i<\chi_i\). When \(\delta_i\geq\chi_i\),
even the first pound costs at least as much as it saves. The normalised
target and two reduced-form parameters are
\begin{equation}
 c_i=
 \begin{cases}
  \overline G_i^{-1}(\delta_i/\chi_i),&0<\delta_i<\chi_i,\\
  0,&\delta_i\geq\chi_i,
 \end{cases}
 \qquad
 h_i=\omega_i c_i,\qquad \nu_i=\psi_i c_i .
 \label{mod:quantile}
\end{equation}
The desired reserve is \(h_i+\nu_i\Delta_i(s)\), where \(h_i\) is the
reserve target under full receipts and \(\nu_i\) its rise per pound of
missed receipts. Write \(\clip(v)=\min\{1,\max\{0,v\}\}\) for clipping to
\([0,1]\), applied coordinatewise for vectors and pointwise for functions.
The early repayment response is
\[
 B_i(s)=\clip\!\left(
       \frac{e_i+z_i(s)-h_i-\nu_i(\bar z_i-z_i(s))}{\bar p_i}\right).
\]
The proof of Proposition~\ref{mod:bestresponses} gives this rule for both
the finite and continuum games. While repayment stays interior, a pound
of missed receipts reduces available cash by one pound and increases
desired reserves by \(\nu_i\) pounds, cutting early payments by
\(1+\nu_i\) pounds. A bank with positive cash pays zero when its desired
reserve is at least as large as its available cash.

\begin{samepage}
\begin{remark}[Date-1 settlement and timing invariance]
Let \(p_i^0=\bar p_i u\) be early repayment,\label{mod:timing-invariance}
\(D_i^{\mathrm{due}}=\bar p_i-p_i^0\)
the deferred principal and \(X_i\geq0\) unrestricted continuation cash at
date 1. The bank pays late principal
\(Q_i^1=\min\{D_i^{\mathrm{due}},a_i(s)-p_i^0+X_i\}\) from restored reserves
and continuation cash, and late interest \(k_iD_i^{\mathrm{due}}\) from the
pledge. Late receipts cannot be reused within the same settlement round.
Total principal repayment and principal loss are
\begin{equation}
 p_i^0+Q_i^1=\min\{\bar p_i,a_i(s)+X_i\},
 \qquad D_i^1=[\bar p_i-a_i(s)-X_i]_+ .
 \label{mod:terminal}
\end{equation}
A pound paid earlier reduces deferred principal and the cash carried into
final settlement by the same amount. At fixed counterparty payments and
continuation cash, both quantities in \eqref{mod:terminal} are therefore
independent of the bank's own early action. A penalty depending only on
terminal principal loss \(D_i^1\) and having finite expectation leaves the
early response intact if the conditional law of \(X_i\) given counterparty
payments is invariant to the
bank's own action. For \(k_i>0\), paying earlier reduces late interest,
an incentive already captured by \(\delta_i\).
\end{remark}
\end{samepage}

\subsection{Directed liability kernels and exact embedding}
\label{mod:representation}

Let \(I=[0,1]\) carry Lebesgue measure. Let \(\bar p:I\to[0,\infty)\)
be the measurable liability profile and \(A=\{x:\bar p(x)>0\}\) the
positive-measure set of active debtor types. Assume
\(0<p_{\min}\leq\bar p\leq p_{\max}<\infty\) on \(A\).
A directed relative-liability kernel is a nonnegative \(P\in L^2(I^2)\) with
\begin{equation}
 \int_I P(x,y)\,dy=1\quad\text{for a.e. }x\in A,
 \qquad P(x,\cdot)=0\quad\text{for a.e. }x\notin A .
 \label{mod:rows}
\end{equation}
The first coordinate indexes the debtor, so \(\bar p(x)P(x,y)\) is the
nominal liability density from \(x\) to \(y\). Types outside \(A\) can receive
payments. For \(s\in\mathcal S_A=\{s\in L^2(I):0\leq s\leq\mathbf1_A
\text{ a.e.}\}\), early cash payments are \(p=\bar p s\). Define
\begin{equation}
 (Kf)(x)=\int_I P(y,x)\bar p(y)f(y)\,dy,
 \qquad z=Ks,\qquad \bar z=K\mathbf1_A .
 \label{mod:inflow}
\end{equation}
The inflow operator \(K:L^2(I)\to L^2(I)\) is positive and Hilbert--Schmidt, with
\(\|K\|_{2\to2}\leq p_{\max}\|P\|_2\). Row normalisation gives
\(\int_I Ks=\int_A\bar p s\). Every unit paid is received by a creditor,
including creditors outside \(A\).

For a finite \(n\)-bank matrix, partition \(I\) into intervals \(I_i\) of
measure \(1/n\). On \(I_i\times I_j\) set \(P_n=n\Pi_{ij}\), and hold the
primitives of bank \(i\) constant on \(I_i\). Write \(K_n^{\mathrm{step}}\)
for the integral operator \eqref{mod:inflow} of this embedded network. For
\(x\in I_i\),
\begin{equation}
 (K_n^{\mathrm{step}}s)(x)=\sum_jL_{ji}\,\langle s\rangle_j,\qquad
 \langle s\rangle_j=n\int_{I_j}s(y)\,dy,\qquad
 \|P_n\|_2^2=\sum_{i,j}\Pi_{ij}^2 .
 \label{mod:embedding}
\end{equation}
On step profiles this is the finite inflow matrix applied to the profile's
cell values. The factor \(n\) preserves per-bank flows. Integrating a step
profile gives a cross-sectional average. If \(\|P_n\|_2\) stays bounded, the
average squared-share sum over all rows is \(O(n^{-1})\). The same bound
holds for the average over active rows if \(|A_n|/n\) stays bounded away from zero.
If the finite network has zero total liabilities, the embedding gives
\(P_n=0\) and the unique profile is zero. The aggregate repayment rate
requires positive total liabilities.

An invertible measure-preserving relabelling \(\pi\) gives
\(P^\pi(x,y)=P(\pi(x),\pi(y))\) and composes every bank characteristic with
\(\pi\). The resulting inflow operator satisfies \(K^\pi(f\circ\pi)=(Kf)\circ\pi\).
Equilibria relabel with their types, and aggregate outcomes are invariant.
Sorting an i.i.d.\ latent sample relabels the realised sample. Each type
can encode a joint sender--receiver position.

\subsection{Best responses and the finite-game correspondence}
\label{mod:responses}

A continuum type \(x\in A\) faces the continuum analogue of
\eqref{mod:netpayoff}, with \(z(x)=(Ks)(x)\) and type-dependent primitives
satisfying the individual bank assumptions above.
The type has measure zero, so the inflow is fixed in its own optimisation.
Assume \(e,\omega,\psi\) are measurable and bounded, \(\chi,\delta\) are finite
and measurable, \((x,t)\mapsto\overline G_x(t)\) is jointly measurable, and the
resulting \(h,\nu\) are essentially bounded on \(A\).
The reserve quantile is measurable by the measurability of the survival
probabilities. Changing the monetary unit scales all cash quantities
together and leaves repayment fractions unchanged. On \(A\) define
\begin{equation}
 g=\frac{e-h-\nu\bar z}{\bar p},\qquad
 Mf=\frac{1+\nu}{\bar p}Kf,\qquad M_0 f=\frac{1}{\bar p}Kf .
 \label{mod:operators}
\end{equation}
The positive compact operators \(M,M_0\) act on \(L^2(A)\), with inputs
extended by zero outside \(A\).

\begin{proposition}[Best responses and embedding]
\label{mod:bestresponses}
For each \(s\in\mathcal S_A\), the unique best response for almost every type in \(A\) is
\begin{equation}
 B(s)=\clip\!\left(
      \frac{e+Ks-h-\nu(\bar z-Ks)}{\bar p}\right)
     =\clip(g+Ms)\quad\text{on }A,\qquad B(s)=0\quad\text{off }A .
 \label{mod:best}
\end{equation}
The map \(B:\mathcal S_A\to\mathcal S_A\) preserves order and satisfies
\begin{equation}
 |B(s)-B(t)|\leq M|s-t|\qquad\text{a.e. on }A.
 \label{mod:domination}
\end{equation}
In the step embedding \eqref{mod:embedding}, its fixed points are exactly
the finite equilibria represented as step profiles, with their order preserved.
\end{proposition}

\subsection{Extremal equilibria and mechanical clearing}
\label{mod:extremal}

A continuum equilibrium is an almost-everywhere fixed point of \(B\).
A finite Nash equilibrium here is a generalised Nash equilibrium because
each bank's feasible actions depend on its counterparties' payments.
The mechanical comparator is
\begin{equation}
 \Phi(s)=\clip\!\left(\frac{e+Ks}{\bar p}\right)
 \quad\text{on }A,\qquad \Phi(s)=0\quad\text{off }A .
 \label{mod:mechanical}
\end{equation}
When \(e\ge0\), \(\Phi\) extends the proportional clearing map of
\citet{eisenbergnoe2001} to directed graphons. Signed \(e\) accommodates
senior external obligations, with available resources given by
\([e+Ks]_+\).

\begin{theorem}[Equilibrium lattice and comparison]
\label{mod:lattice}
Strategic equilibria form a nonempty complete lattice. Iteration of \(B\)
from \(\mathbf1_A\) and zero gives the greatest and least equilibria,
\(s^+\) and \(s^-\), respectively, almost everywhere and in \(L^2\).
The same holds for \(\Phi\), with extremal equilibria \(s_\Phi^+,s_\Phi^-\),
and
\begin{equation}
 s^+\leq s_\Phi^+,\qquad s^-\leq s_\Phi^- .
 \label{mod:comparison}
\end{equation}
If \(\delta\geq\chi\) almost everywhere on \(A\), then \(B=\Phi\).
\end{theorem}

While type \(x\)'s best response remains interior, one extra unit of receipts
lowers its desired reserves by \(\nu(x)=\psi(x)c(x)\) units. Each unit of
missed receipts raises its expected funding need by
\(\psi(x)\mathbb E[Y_x]\), which can differ from the reserve response. Writing
the aggregate early repayment rate as
\begin{equation}
 \operatorname{RR}(s)=\frac{\int_A\bar p(x)s(x)\,dx}{\int_A\bar p(x)\,dx},
 \label{mod:rr}
\end{equation}
the excess shortfall for either pair of equilibrium extrema
\((s^G,s_\Phi)\in\{(s^-,s_\Phi^-),(s^+,s_\Phi^+)\}\) decomposes as
\begin{align}
 \operatorname{RR}(s_\Phi)-\operatorname{RR}(s^G)
 &=\big[\operatorname{RR}(\Phi(s^G))-\operatorname{RR}(s^G)\big]\notag\\
 &\quad+\big[\operatorname{RR}(s_\Phi)-\operatorname{RR}(\Phi(s^G))\big].
 \label{mod:decomp}
\end{align}
Since \(s^G=B(s^G)\leq\Phi(s^G)\leq\Phi(s_\Phi)=s_\Phi\), both terms are
nonnegative. They measure the payments banks withhold from payable
resources and the capacity they lose when counterparties pay less.
Equation~\eqref{mod:terminal} separates early non-payment from terminal
principal loss. For a common continuation cash schedule, a higher early
repayment profile weakly lowers terminal principal losses through its
effect on receipts.

\subsection{Uniqueness}
\label{mod:uniqueness}

Let \(T_{AA}f(x)=\int_A P(x,y)f(y)\,dy\) on \(L^2(A)\), and let
\(\ell_{\mathrm{out}}(x)=\int_{I\setminus A} P(x,y)\,dy\) be the share owed by an active
debtor to pure creditors. Write \(\rho\) for spectral radius.

\begin{proposition}[Spectral conditions]
\label{mod:spectral}
If \(\rho(M)<1\), the strategic equilibrium is unique, and iteration from any
strategy converges to it in \(L^2\). Furthermore,
\begin{equation}
 \rho(M_0)=\rho(T_{AA})\leq 1-\operatorname*{ess\,inf}_{x\in A}\ell_{\mathrm{out}}(x).
 \label{mod:leakage}
\end{equation}
The condition \((1+\|\nu\|_\infty)\rho(T_{AA})<1\) suffices for strategic
uniqueness. Uniformly positive leakage implies mechanical uniqueness.
\end{proposition}

The order and clipping of the responses give the two monotone endpoint
iterations. The spectral condition controls their separation through
successive powers of \(M\). With the liability kernel and other primitives
fixed, if \(\delta_m\) increases and \(c_m\downarrow0\) almost everywhere, both
strategic extrema increase to fixed points of \(\Phi\). When mechanical
clearing is unique, they converge to that profile in \(L^2\).

\section{From balance sheets to sampled networks}
\label{samp:sampling}

With incomplete bilateral data, balance-sheet totals define the set of
feasible point networks. A reconstruction selects a matrix from that set.
A fitted graphon and a sampling law then specify how bilateral exposures
vary across draws.

\subsection{What the balance-sheet totals determine}
\label{samp:margins}

Suppose that the bilateral matrix is unobserved and that each bank's total
nominal liabilities \(\bar p_i\) and total contractual claims \(\bar z_i\)
are known. Write \(\bar p=(\bar p_i)_{i=1}^n\) and
\(\bar z=(\bar z_i)_{i=1}^n\) for these vectors. A reconstruction selects
a matrix from the feasible set
\begin{equation}
 \mathcal L(\bar p,\bar z)
 =\{L\in[0,\infty)^{n\times n}:L\mathbf 1=\bar p,\;
 L^\top\mathbf 1=\bar z,\; L_{ii}=0\text{ for all }i\}.
 \label{samp:feasible-margins}
\end{equation}
Assume this set is nonempty and hold the liquidity and reserve parameters
\(e_i,h_i,\nu_i\) fixed across its elements.

\begin{proposition}[Corner equilibria at fixed balance-sheet totals]
\label{samp:fixed-margin-corners}
For every matrix in \(\mathcal L(\bar p,\bar z)\), zero early payment is an
equilibrium if and only if \eqref{samp:zero-margin} holds for every \(i\in A_n\).
Full early payment on \(A_n\) is an equilibrium if and only if
\eqref{samp:full-margin} holds for every \(i\in A_n\).
\begin{subequations}
\label{samp:margin-tests}
\begin{align}
 e_i &\leq h_i+\nu_i\bar z_i, \label{samp:zero-margin}\\
 e_i+\bar z_i &\geq \bar p_i+h_i. \label{samp:full-margin}
\end{align}
\end{subequations}
Consequently, all same-margins reconstructions agree on the existence of
each corner, conditional on the bank parameters.
\end{proposition}

At zero payment, bank \(i\) receives zero and its desired reserve is
\(h_i+\nu_i\bar z_i\). At full payment, it receives \(\bar z_i\) and its
desired reserve is \(h_i\). Substitution into \eqref{mod:best} gives the two
tests. At intermediate profiles, realised receipts are
\(\sum_j L_{ji}s_j\), which can vary across reconstructions even when both
totals are fixed. Section~\ref{ex:matched-marginals} gives an exact example
in which this changes the least equilibrium.

\subsection{From a point reconstruction to a sampling kernel}
\label{subsec:estimation-bridge}

Given an observed \(L\) or a matrix reconstructed from
\eqref{samp:feasible-margins}, form the relative shares \(\Pi\) as in
Section~\ref{mod:representation}. A maximum-entropy allocation over eligible
links is one approach to reconstruction from balance-sheet data
\citep{anandcraigvonpeter2015}. A fitness-based support model can
instead propose links, followed by fitting positive weights to both margins
when the support is feasible. See \citet{gandyveraart2017} for a Bayesian
approach to network reconstruction. To fit a directed surface, assign
each bank a sender and a receiver coordinate \((\phi_i,\varphi_i)\). The
left and right singular vectors of \(\Pi\) provide one assignment. The
coordinates and the balance-sheet marks are retained jointly for each bank.

One specification fits a nonnegative affinity \(W\) to off-diagonal
densities \(n\Pi_{ij}\). Let \(\phi(x)\) and \(\varphi(x)\) denote the sender
and receiver coordinates of type \(x\), with values \(\phi_i,\varphi_i\) at
an observed bank's assigned type \(x_i\). Set \(Q(x,y)=W(\phi(x),\varphi(y))\).
With positive normalising mass on the active rows, define
\[
 P(x,y)=\frac{Q(x,y)}{\int_I Q(x,v)\,\mathrm dv}\quad(x\in A),\qquad
 P(x,\cdot)=0\quad(x\notin A).
\]
A bounded \(Q\) with normalising mass bounded away from zero gives a bounded
\(P\). This separates the fitted scores \(Q\), the relative-liability density
\(P\), and the sharpened population target \(P_\tau\) in
\eqref{samp:target}.

\subsection{Sparsemax--Dirichlet network generation}
\label{subsec:sparsemax-sampler}

Fix a bounded, row-normalised liability kernel \(P\), a sharpening parameter
\(\tau\in(0,1]\) and a row concentration \(\gamma\in(0,\infty]\).
Sparsemax \citep{martinsastudillo2016} allocates the debtor's budget across
a selected set of creditors by simplex projection. The Dirichlet draw varies
the positive weights around their projected means.

\begin{enumerate}
\setlength{\itemsep}{0.2em}
\item \textit{Choose banks and marks.} For a synthetic economy, draw
\(x_1,\ldots,x_n\) independently and uniformly on \([0,1]\), with \(n\ge2\),
and set \((\bar p_i,e_i,h_i,\nu_i)=(\bar p,e,h,\nu)(x_i)\) using fixed bounded
Borel representatives of the kernel and marks. An observed-bank
rewiring exercise instead holds the bank set, assigned positions and marks
fixed (Section~\ref{subsec:estimation-bridge}). The convergence results below
use the synthetic design.
\item \textit{Mask and normalise scores.} For each active debtor
\(i\in A_n\), exclude \(i\) from its creditor row and set
\(q_{ij}=P(x_i,x_j)/\sum_{k\ne i}P(x_i,x_k)\) for \(j\ne i\).
Use \(q_{ij}=1/(n-1)\) if the denominator is zero. The type-sampling bound
makes such rows rare under the synthetic design. For a fixed bank set,
report any fallback rows, where all off-diagonal positions become eligible.
\item \textit{Project and identify the support.} Project the vector
\((q_{ij}/\tau)_{j\ne i}\) onto the simplex of \(n-1\) nonnegative
shares summing to one. Let
\(q_{i(1)}\ge\cdots\ge q_{i(n-1)}\) be the sorted scores within the row. Then
\[
\begin{gathered}
 m_i=\max\left\{k\in\{1,\ldots,n-1\}:\sum_{\ell=1}^k(q_{i(\ell)}-q_{i(k)})<\tau\right\},
 \qquad
 t_i=\frac{\tau^{-1}\sum_{\ell=1}^{m_i}q_{i(\ell)}-1}{m_i},\\
 w_{ij}=[q_{ij}/\tau-t_i]_+ \ (j\ne i),\qquad
 w_{ii}=0,\qquad \sum_{j\ne i}w_{ij}=1 .
\end{gathered}
\]
The strict inequality selects the positive coordinates. A score
\(q_{ij}=\tau t_i\) receives zero weight. Since
\(\sum_{j\ne i}[q_{ij}-\tau t_i]_+=\tau\), the supports
\(S_i=\{j:w_{ij}>0\}\) can only shrink as \(\tau\) decreases, with \(m_i=|S_i|\).
At \(\tau=1\), \(w_{ij}=q_{ij}\) for \(j\ne i\). On inactive rows,
\(w_{ij}=0\) for all \(j\), \(S_i=\varnothing\) and \(m_i=0\).
\item \textit{Realise the weights.} Conditional on the types, draw independently
across active rows for finite \(\gamma\). In the deterministic allocation
mode, \(\gamma=\infty\), use the projected weights directly. Each supported
coordinate is positive almost surely in both modes, so \(\gamma\) changes
the allocation on the support and leaves its exact zeros fixed.
\item \textit{Recover nominal liabilities.} Multiply each row by its debtor's
nominal obligations. The two allocation modes give
\begin{equation}
\begin{gathered}
 (\Pi_{ij})_{j\in S_i}\ \sim\
 \operatorname{Dirichlet}\big((\gamma w_{ij})_{j\in S_i}\big)
 \quad(\gamma<\infty),\\
 (\Pi_{ij})_{j\in S_i}=(w_{ij})_{j\in S_i}\quad(\gamma=\infty),\qquad
 \Pi_{ij}=0\ \text{if }j\notin S_i,\qquad L_{ij}=\bar p_i\Pi_{ij}.
\end{gathered}
 \label{samp:sampler}
\end{equation}
Set inactive rows to zero. A singleton support gives unit weight on its one
creditor in both modes. Thus \(L_{ii}=0\) and \(\sum_j L_{ij}=\bar p_i\) exactly.
\end{enumerate}

\paragraph{The scale of the projection.}
Direct projection of \((P(x_i,x_j)/\tau)_{j\ne i}\) uses an implicit row temperature
\(\tau/\sum_{j\ne i} P(x_i,x_j)\) on the normalised scores \((q_{ij})_{j\ne i}\). That
temperature is of order \(1/n\) under the synthetic design at fixed \(\tau\),
since the empirical row average converges uniformly to one on active rows. Row normalisation gives a common
scale for sharpening and for the limit below.

\paragraph{Population target.}
For almost every \(x\in A\), let \(\theta_\tau(x)\ge0\) solve
\begin{equation}
 \int_0^1 [P(x,y)/\tau-\theta_\tau(x)]_+\,\mathrm dy=1,\qquad
 P_\tau(x,y)=[P(x,y)/\tau-\theta_\tau(x)]_+ .
 \label{samp:target}
\end{equation}
Set \(P_\tau=0\) on inactive rows. If \(P\le P_{\max}\), the left side decreases
continuously and strictly from \(1/\tau\) while positive, and is zero by
\(P_{\max}/\tau\). Taking the infimum of thresholds with integral at most one
gives measurability. The target has unit row mass and \(P_\tau\le P_{\max}/\tau\).
In particular \(P_1=P\). For \(\tau<1\), equality holds almost everywhere
exactly when almost every active row is constant on its positive support.
On that support equality requires
\((\tau^{-1}-1)P(x,y)=\theta_\tau(x)\), which also suffices for such a row.
Taking \(P_{\max}(1/\tau-1)\) as a threshold makes the excess at most \(P\),
so \(\|P_\tau-P\|_\infty\le P_{\max}(1/\tau-1)\).
Write \(K_\tau\) for the receipt operator \eqref{mod:inflow} with kernel
\(P_\tau\). The target contractual claims are \(\bar z_\tau=K_\tau\mathbf1_A\).

The simplex weights must be rescaled to compare them with a density.
Put \(N=n-1\) and \(D_{n,i}=N^{-1}\sum_{j\ne i}P(x_i,x_j)\).
For active \(i\) with positive score mass and \(j\ne i\),
\[
 Nw_{ij}=[P(x_i,x_j)/(\tau D_{n,i})-Nt_i]_+ .
\]
For fixed \(\tau\), Lemma~\ref{samp:inflows} gives the almost-sure errors
\(Nt_i-\theta_\tau(x_i)=O(\sqrt{\log n/n})\) uniformly over active rows and
\(w_{ij}-P_\tau(x_i,x_j)/N=O(\sqrt{\log n/n}/N)\) uniformly over their
off-diagonal entries.
The same target governs actual and contractual receipts. Changing \(\tau\)
can change the contractual cash due to creditors even when the support is
unchanged.

\paragraph{Support and exposure concentration.}
Since \(P_\tau\le P_{\max}/\tau\) and each active row integrates to one, its
positive support has measure at least \(\tau/P_{\max}\). At fixed \(\tau\)
the sampled active rows likewise have order \(n\) positive positions with high
probability. The squared-share sum \(\sum_j\Pi_{ij}^2\) can still decay more
slowly than \(1/n\) when \(\gamma_n\) grows sublinearly. Smaller values of
\(\tau\) can remove weak links in a finite sample. For rank-one power-law
scores \(Q(x,y)=(xy)^{-\alpha}\), \(0<\alpha<1\), regularising the creditor
factor to \((y+\epsilon)^{-\alpha}\) with fixed \(\epsilon>0\) gives a bounded
row-normalised kernel in the regime of Section~\ref{samp:approximation}.

\subsection{What the sampling law preserves}
\label{samp:accounting}

For an active row, conditional on \(\boldsymbol{x}=(x_1,\ldots,x_n)\), the
Dirichlet moments give
\begin{equation}
\begin{gathered}
 \mathbb E[\Pi_{ij}\mid \boldsymbol{x}]=w_{ij},\qquad
 \operatorname{Var}(\Pi_{ij}\mid \boldsymbol{x})=\frac{w_{ij}(1-w_{ij})}{\gamma+1},\\
 \operatorname{Cov}(\Pi_{ij},\Pi_{ik}\mid \boldsymbol{x})=
 -\frac{w_{ij}w_{ik}}{\gamma+1}\quad(j\ne k),\\
 \mathbb E\!\left[\sum_j\Pi_{ij}^{2}\,\middle|\, \boldsymbol{x}\right]=
 H_i^w+\frac{1-H_i^w}{\gamma+1},\qquad H_i^w=\sum_j w_{ij}^{2}.
\end{gathered}
\label{samp:dirichlet-moments}
\end{equation}
The realised nominal inflow \(\bar z_i^{(n)}=\sum_{k\ne i}\bar p_k\Pi_{ki}\) satisfies
\begin{equation}
 \mathbb E[\bar z_i^{(n)}\mid \boldsymbol{x}]=\sum_{k\ne i}\bar p_k w_{ki},\qquad
 \operatorname{Var}(\bar z_i^{(n)}\mid \boldsymbol{x})
 =\frac{\sum_{k\ne i}\bar p_k^2 w_{ki}(1-w_{ki})}{\gamma+1}.
 \label{samp:nominal-moments}
\end{equation}
Use \(1/(\gamma+1)=0\) in the deterministic mode. The finite game uses
the realised \(\bar z_i^{(n)}\) as the nominal amount due when computing
missed receipts. The sampler fixes debtor totals. Creditor totals are
determined by the generated shares.

\paragraph{Calibrating support and concentration.}
Where observed creditor counts are available, choose \(\tau\) by matching
their distribution, or its mean, against the counts \(m_i\) on the same
bank set.
Given \(\tau\), let \(\bar H\) be the average observed squared share sum
over active debtors and \(\bar H^w\) the average for their projected rows. A finite
positive \(\gamma\) matches this moment when \(\bar H^w<\bar H<1\), with
\(\gamma=(1-\bar H^w)/(\bar H-\bar H^w)-1\). The deterministic mode matches
\(\bar H=\bar H^w\). Report the selected parameters alongside formal creditor
counts, effective counts \(1/\sum_j\Pi_{ij}^2\), and the discrepancies in
creditor totals. Conditional on the types and \(\tau\), \(\gamma\) changes
the dispersion around row \((w_{ij})_j\). Clipping and feedback can carry that
dispersion through to mean repayment.

\begin{table}[tbp]
 \centering
 \caption{Finite, continuum and sampled repayment economies}
 \label{samp:map}
 \small
 \begin{tabularx}{\linewidth}{@{}>{\raggedright\arraybackslash}p{0.185\linewidth}*{3}{>{\raggedright\arraybackslash}X}@{}}
 \toprule
 Object & Finite network & Continuum economy & Sparsemax-sampled network \\
 \midrule
 Institution & \(i=1,\ldots,n\) & Type \(x\in I\) &
 New draws \(x_i\) or assigned positions \\
 Balance sheets and reserves & \((\bar p_i,e_i,h_i,\nu_i)\) &
 \((\bar p,e,h,\nu)(x)\) & Evaluate at \(x_i\) or hold the observed marks fixed \\
 Relative liabilities & \(\Pi_{ij}\), with \(\Pi_{ii}=0\) &
 \(P\), or \(P_\tau\) after sharpening &
 Centres \(w_{ij}\), realised shares \(\Pi_{ij}\) from \eqref{samp:sampler} \\
 Contractual claims & \(\bar z_i=\sum_j L_{ji}\) &
 \(\bar z_\tau=K_\tau\mathbf1_A\) under \(P_\tau\) &
 \(\bar z_i^{(n)}=\sum_j\bar p_j\Pi_{ji}\) \\
 Receipts & \(\sum_j L_{ji}s_j\) &
 \(K_\tau s\) under \(P_\tau\) &
 \(\sum_j\bar p_j\Pi_{ji}s_j\) with realised shares \\
 Repayment equilibrium & Fixed point of the finite response &
 Fixed point of \(B_\tau\) &
 Finite fixed point on the generated liabilities \\
 \bottomrule
 \end{tabularx}
\end{table}

Table~\ref{samp:map} maps the same repayment quantities across the three
economies. The approximation results compare their equilibria after
aligning the sampled institutions with their continuum types.

\section{Which equilibria survive sampling?}
\label{samp:approximation}
\label{pers:section}

For the sharpened kernel \(P_\tau\), define the intercept and feedback
operators on \(A\) by
\[
 g_\tau=\frac{e-h-\nu\bar z_\tau}{\bar p},\qquad
 M_\tau f=\frac{1+\nu}{\bar p}K_\tau f,\qquad
 M_{0,\tau}f=\frac{1}{\bar p}K_\tau f.
\]
Write \(\mathcal U_\tau(f)=g_\tau+M_\tau f\) for the affine index and
\(B_\tau(f)=\clip(\mathcal U_\tau(f))\) for the continuum response.
Repayment profiles and response indices are zero off their respective
active sets. For fixed Borel representatives of continuum functions,
let \(R_nf=(f(x_i))_{i=1}^n\), extending functions defined only on \(A\) by
zero. For a set of indices \(S\), write \(u_S=(u_i)_{i\in S}\).
The sampled receipt matrix \(K_n\) and nominal inflows are
\[
 (K_nu)_i=\sum_{k\ne i}\bar p_k\Pi_{ki}u_k,\qquad
 \bar z_i^{(n)}=(K_nR_n\mathbf 1_A)_i \quad (1\le i\le n).
\]
On \(A_n\), set
\[
 g_{n,i}=\frac{e_i-h_i-\nu_i\bar z_i^{(n)}}{\bar p_i},\qquad
 (M_nu)_i=\frac{1+\nu_i}{\bar p_i}(K_nu)_i,\qquad
 (M_{0,n}u)_i=\frac{1}{\bar p_i}(K_nu)_i,
\]
and set their entries to zero off \(A_n\). The sampled affine index is
\(\mathcal U_n(u)=g_n+M_nu\), and the sampled best response is
\begin{equation}
 B_n(u)=\clip(\mathcal U_n(u)).
 \label{samp:sample-response}
\end{equation}
Here \(P\le P_{\max}<\infty\), \(0<|A|\le1\),
\(0<p_{\min}\le\bar p\le p_{\max}<\infty\) on \(A\), and \(\nu\) is bounded.
Use the synthetic design in \eqref{samp:sampler} with independently drawn types
within each sample, fixed \(\tau\) and concentration \(\gamma_n\) unless stated
otherwise. Since \(\Pr(A_n=\varnothing)=(1-|A|)^n\), almost surely \(A_n\) is
nonempty for all sufficiently large \(n\). Set
\(r_n=\sqrt{\log n/n}+\sqrt{\log n/\gamma_n}\), with
\(1/\gamma_n=0\) in the deterministic case.

\subsection{Convergence of equilibrium sets}
\label{samp:ne-section}
\label{samp:common-space}

Order the sampled types as \(x_{(1)}\le\cdots\le x_{(n)}\). Let \(u_{(i)}\)
be the coordinate of \(u\) attached to \(x_{(i)}\). Its sorted step profile
\(\widetilde u_n\) equals \(u_{(i)}\) on \(I_i=((i-1)/n,i/n]\). Write
\(T_n(x)=x_{(i)}\) on \(I_i\). The sorted step profile of \(R_nf\) is then
\(f\circ T_n\) almost everywhere.
Let \(B_n^{\mathrm{step}}\) be the response associated with this sorted step
embedding. By
Proposition~\ref{mod:bestresponses}, finite equilibria and their step profiles
are equivalent descriptions of the same game.

Let \(\mathcal E_\tau\subset L^2([0,1])\) be the Nash equilibrium set of the
continuum game with kernel \(P_\tau\), and let \(\widetilde{\mathcal E}_n\)
be the set of sorted step profiles of the finite Nash equilibria. Both sets are
nonempty by Theorem~\ref{mod:lattice} and the exact finite-game embedding.
When \(A_n=\varnothing\), \(\widetilde{\mathcal E}_n=\{0\}\).

\begin{theorem}[Finite-network and graphon Nash equilibria]
\label{samp:ne-limit}
For fixed \(\tau\in(0,1]\), suppose \(P\) is bounded and the sampling law
\eqref{samp:sampler} has \(\gamma_n\to\infty\). Almost surely,
\begin{equation}
 \sup_{v\in\widetilde{\mathcal E}_n}
   \inf_{s\in\mathcal E_\tau}\|v-s\|_{L^2}
 \longrightarrow 0 .
 \label{samp:ne-set}
\end{equation}
In particular, if the graphon Nash equilibrium \(s_\tau\) is unique, then
\begin{equation}
 \sup_{v\in\widetilde{\mathcal E}_n}\|v-s_\tau\|_{L^2}
 \longrightarrow 0
 \qquad\text{almost surely}.
 \label{samp:ne-unique}
\end{equation}
\end{theorem}

The proof controls the sampled liability kernel in cut norm and uses
compactness of the continuum response.

The limit concerns \(P_\tau\), with \(P_\tau=P\) for \(\tau=1\).
Under the same concentration condition, the \(L^2\) conclusions target \(P\)
when the deterministic sequence \(\tau_n\) tends to one. The mean-weight
bounds are uniform over temperatures bounded away from zero, and the
threshold bound following \eqref{samp:target} gives \(\|P_{\tau_n}-P\|_\infty\to0\).
Theorem~\ref{pers:noncontractive} gives the reverse approximation at each
graphon equilibrium with strict repayment margins and an invertible
interior equilibrium equation.

\subsection{Exact persistence of a continuum branch}
\label{pers:noncontractive-section}

For an equilibrium \(s^\star\) under \(P_\tau\), partition \(A\) into
\(Z=\{x\in A:s^\star(x)=0\}\), \(J=\{x\in A:0<s^\star(x)<1\}\) and
\(F=\{x\in A:s^\star(x)=1\}\). The sampled sets \(Z_n,J_n,F_n\) contain the
indices whose types lie in \(Z,J,F\), respectively. Let \(M_{\tau,JJ}\) be
\(M_\tau\) restricted to outputs on \(J\) and inputs in \(L^2(J)\) extended by
zero. Say the equilibrium has strict margins if some \(\eta>0\) satisfies, almost everywhere,
\begin{equation}
 \mathcal U_\tau(s^\star)\le -\eta\ \text{on }Z,\qquad
 \eta\le s^\star\le 1-\eta\ \text{on }J,\qquad
 \mathcal U_\tau(s^\star)\ge 1+\eta\ \text{on }F .
 \label{pers:margins}
\end{equation}
Under these margins, only banks with interior choices pass small changes in
receipts through to repayment locally.

\begin{theorem}[Exact equilibria near a nondegenerate branch]
\label{pers:noncontractive}
Suppose the sampling conditions of Theorem~\ref{samp:ne-limit} hold, including
\(\gamma_n\to\infty\). Let \(s^\star\) be a continuum equilibrium with the
strict margins in \eqref{pers:margins}. If \(1\notin\sigma(M_{\tau,JJ})\),
where \(\sigma\) denotes the spectrum on \(L^2(J)\), then almost surely there exist exact finite equilibria whose
sorted step profiles converge to \(s^\star\) in \(L^2\). The spectral
condition is vacuous for a null interior set.
\end{theorem}

The proof treats strict margins as local stability of the clipping regions. The
invertible interior equation and a fixed-point argument then give an exact
sampled equilibrium nearby.

If every continuum equilibrium satisfies the hypotheses, the equilibrium sets
converge in Hausdorff distance in \(L^2\). The direction from finite to
continuum equilibria is \eqref{samp:ne-set}. For the reverse direction,
take a deterministic finite \(1/k\)-net of continuum equilibria for each
integer \(k\ge1\), using compactness. Intersect the probability-one events
for all net points. On that event, for each \(k\) the reverse distance is
at most \(2/k\) for all sufficiently large \(n\).

\subsection{Survival of a freeze and full settlement}
\label{pers:finite-sample}

The corner tests in Proposition~\ref{samp:fixed-margin-corners} also apply
almost everywhere to \(P_\tau\). To protect both tests against sampling error,
suppose that opening liquidity stays a strict distance from each threshold.
Specifically, for some \(\eta>0\) require
\begin{equation}
 \bar p-\bar z_\tau+h+\eta\bar p\ \le e\ \le\
 h+\nu\bar z_\tau-\eta\bar p
 \quad\text{a.e. on }A.
 \label{pers:strict-corners}
\end{equation}
The difference between the unbuffered upper and lower thresholds is \((1+\nu)\bar z_\tau-\bar p\).
At full payment all due inflows arrive, so the confidence coefficient
\(\nu\) has no effect on the lower threshold. In the canonical sampler,
contractual claims can shift even though debtor totals are preserved.
The strict margins bound the amount of such error the two equilibria can
withstand.

When \(\rho(M_{0,\tau})<1\), set
\[
 \zeta_0=\sum_{k\ge 0}M_{0,\tau}^k\mathbf 1_A,\qquad V_0=\|\zeta_0\|_\infty,
\]
with \(\zeta_0=0\) off \(A\). The bounded kernel and the spectral-radius bound give \(V_0<\infty\),
\(\zeta_0\ge1\) and \(M_{0,\tau}\zeta_0=\zeta_0-\mathbf 1_A\) on \(A\).

\begin{proposition}[Finite-sample coexistence and mechanical uniqueness]
\label{pers:finite-bound}
Suppose the strict corner conditions \eqref{pers:strict-corners} hold and
\(\rho(M_{0,\tau})<1\). For a sample of size \(n\ge2\)
with concentration \(\gamma\in(0,\infty]\), fix a failure probability \(\alpha\in(0,1)\)
and evaluate the constants in Lemma~\ref{samp:inflows} at
\(\ell=\log(14n/\alpha)\). If \(\epsilon\le1/2\), \(a_\gamma<1\), and
\begin{equation}
 E\le p_{\min}\min\left\{\frac{\eta}{\max\{1,\|\nu\|_\infty\}},
                       \frac{1}{2V_0}\right\},
 \label{pers:finite-condition}
\end{equation}
then with probability at least \(1-\alpha\), all sampled active banks pay zero
in the least strategic equilibrium and one in the greatest, and the unique
mechanical clearing vector is full early payment. The probability that these
conclusions hold and \(A_n\ne\varnothing\) is at least
\(1-\alpha-(1-|A|)^n\). On this event, zero and full early payment on \(A_n\)
are distinct strategic equilibria.
\end{proposition}

\begin{corollary}[Eventual survival of full settlement and a freeze]
\label{pers:corners}
Suppose \(\gamma_n/\log n\to\infty\) and the strict continuum corner
conditions \eqref{pers:strict-corners} hold for \(P_\tau\). Almost surely, the
sampled least and greatest strategic equilibria are zero and one on the
active banks for all sufficiently large \(n\). If \(\rho(M_{0,\tau})<1\), the sampled
mechanical clearing vector is eventually almost surely unique and full.
\end{corollary}

The strict margins matter at the endpoints of a coordination region. For
example, a closed core with \(\nu=0\), \(\bar z_\tau=\bar p\) and \(e=h\) has every
constant repayment fraction as an equilibrium. A uniform decrease in $e$
makes zero the sole equilibrium in the one-block homogeneous example.
This discontinuity can occur even though the shift in the best-response
operator is arbitrarily small.

\subsection{Uniform accuracy at sampled banks}
\label{samp:uniform-section}

When \(\gamma_n/\log n\to\infty\), the receipt approximation
in Appendix~\ref{samp:proof-appendix} gives errors of order \(r_n\) for fixed
profiles.

\begin{theorem}[Nash equilibrium convergence under contraction]
\label{samp:convergence}
Suppose \(\rho_{L^2(A)}(M_\tau)<1\) and \(\gamma_n/\log n\to\infty\). Write \(s_\tau\) for the unique equilibrium of the sharpened kernel \(P_\tau\). There exist constants \(C_R,C_M<\infty\), \(\beta<1\) and a bounded measurable \(\zeta\ge 1\) on \(A\) such that almost surely, for all sufficiently large \(n\), the sampled network has a unique Nash equilibrium \(s^{(n)}\) and
\begin{equation}
 \max_{i\in A_n}|s_i^{(n)}-s_\tau(x_i)|
 \le \frac{\|\zeta\|_\infty C_Rr_n}{1-\beta-C_Mr_n}.
 \label{samp:rate}
\end{equation}
In particular, the sorted step profile of the finite Nash equilibrium converges to the graphon Nash equilibrium:
\begin{equation}
 \|\widetilde s_n-s_\tau\|_{L^2}\longrightarrow0
 \qquad\text{almost surely}.
 \label{samp:l2-limit}
\end{equation}
If \(s_\tau\) is piecewise Lipschitz as in Lemma~\ref{samp:step}, the left-hand
side of \eqref{samp:l2-limit} is \(O(r_n+(\log n/n)^{1/4})\) almost surely.
\end{theorem}

\begin{corollary}[Aggregate repayment and mechanical clearing]
\label{samp:aggregate}
Under Theorem~\ref{samp:convergence}, the liability-weighted early repayment rate of the sample converges almost surely to
\[
 \mathrm{RR}_\tau=\frac{\int_0^1 \bar p(x)s_\tau(x)\,\mathrm dx}{\int_0^1\bar p(x)\,\mathrm dx}.
\]
The analogous convergence holds for the mechanical repayment rate on the same liability network, and for the difference between mechanical and strategic rates.
\end{corollary}

\begin{proposition}[Bounds for all sampled equilibria]
\label{pers:bracket}
Suppose \(\gamma_n/\log n\to\infty\). Almost surely, for every
$\varepsilon>0$ and all sufficiently large $n$, every sampled equilibrium $v$
satisfies
\[
 R_ns^- -\varepsilon\mathbf 1_{A_n} \ \le v\le\
 R_ns^+ +\varepsilon\mathbf 1_{A_n} ,
\]
where \(\mathbf 1_{A_n}\) is the indicator vector of \(A_n\) and
\(s^-,s^+\) are the continuum extremal equilibria under \(P_\tau\). In particular, if the
continuum equilibrium is unique, every sampled equilibrium converges uniformly to
its evaluations almost surely.
\end{proposition}

\begin{theorem}[Persistence with several equilibria]
\label{pers:branch}
Suppose \(\gamma_n/\log n\to\infty\). Let \(s^\star\) be a continuum equilibrium
under \(P_\tau\) satisfying \eqref{pers:margins} and \(\rho(M_{\tau,JJ})<1\),
with the spectral condition vacuous for a null interior set. There is a
deterministic radius \(r>0\), independent of \(n\), such that almost surely
for all sufficiently large \(n\) the sampled game has exactly one equilibrium
within maximum-norm distance \(r\) of \(R_ns^\star\). This equilibrium has the sampled
partition \((Z_n,J_n,F_n)\) and is within \(O(r_n)\) of \(R_ns^\star\). Its sorted step
profile converges to $s^\star$ in $L^2$ almost surely. If $s^\star$ is the
continuum least or greatest equilibrium, the finite equilibrium in the
neighbourhood is eventually almost surely the corresponding finite extremal equilibrium.
\end{theorem}

\section{When aggregation fails}
\label{ex:fixed-gamma}

Consider three equal-measure blocks \(U,V,D\). Wholesale counterparties in
\(U\) owe dealer banks in \(V\), which owe downstream cash clients in \(D\).
Each debtor owes one unit, and the clients have no outgoing obligations
in the modelled network.
The two nonzero kernel rectangles have constant density three, so Sparsemax
leaves this kernel unchanged. Set \(e_U=e_V=1\), \(h_U=\nu_U=\nu_V=0\), and
\(h_V=1\). Each upstream debtor pays one. Each dealer in \(V\) pays its
receipts capped at one. In the continuum, dealer receipts equal one and
the acyclic game has a unique full-payment equilibrium. Since every
debtor has \(e_i=\bar p_i=1\) in the finite game, mechanical clearing
uniquely pays in full and terminal principal loss is zero at every
feasible early-payment profile, even with \(X_i=0\).

\begin{theorem}[Exact aggregate repayment with fixed allocation concentration]
\label{ex:fixed-limit}
Use the sampler in \eqref{samp:sampler} with fixed \(\tau\in(0,1]\) and
\(\gamma>0\). Let \(Z_\gamma\) be the sum of the points of a Poisson random
measure on \((0,1)\) with intensity
\(\mu_\gamma(dx)=\gamma x^{-1}(1-x)^{\gamma-1}\,dx\). Its Laplace transform is
\begin{equation}
 \mathbb E[e^{-tZ_\gamma}]
 =\exp\!\left[-\gamma\int_0^1
 (1-e^{-tx})\frac{(1-x)^{\gamma-1}}{x}\,dx\right],\qquad t\geq0.
 \label{ex:fixed-laplace}
\end{equation}
When \(A_n\) is nonempty, let \(\mathrm{RR}_n\) be the average over \(i\in A_n\)
of \(s_i^{(n)}\) at any sampled equilibrium. This is the liability-weighted
repayment rate because each debtor owes one. Almost surely,
\begin{equation}
 \mathrm{RR}_n\longrightarrow 1-D_\gamma,\qquad
 D_\gamma=\tfrac12\mathbb E[(1-Z_\gamma)_+]>0.
 \label{ex:fixed-gap}
\end{equation}
The receipt law has mean one and variance \(1/(\gamma+1)\), and
\begin{equation}
 \sqrt{\gamma}\,D_\gamma\longrightarrow\frac{1}{2\sqrt{2\pi}}
 \quad\text{as }\gamma\to\infty .
 \label{ex:fixed-large}
\end{equation}
\end{theorem}

The receipt law is the total mass of a beta process with unit base mass and
concentration \(\gamma\) \citep{thibauxjordan2007}. Upstream counterparties pay
in full, yet individual dealers receive different amounts. Low receipts
can leave a dealer below its payment cap, while extra receipts at a dealer
already at the cap do not raise aggregate payments. Blocks \(U\) and \(V\)
each account for half of active
obligations in the limit, and only \(V\) can fall short. The mean shortfall
in \(V\) therefore enters \eqref{ex:fixed-gap} with weight \(1/2\).

The sampler fixes each debtor's outgoing liabilities and allows nominal
claims to vary. In this example every upstream debtor pays in full, so
each dealer's realised receipts equal its nominal claims.
Holding those claims fixed removes this source of receipt dispersion
across draws. In particular, unit incoming claims give full repayment at
each dealer. The fixed-\(\gamma\) result thus measures the
effect of exposure allocation in the canonical sampler. Comparisons
that condition on both balance-sheet totals are discussed separately
in Section~\ref{ex:blocks}.

\begin{proposition}[More diffuse allocations reduce the limiting shortfall]
\label{ex:fixed-order}
The gap \(D_\gamma\) is strictly decreasing in \(\gamma>0\). As
\(\gamma\downarrow0\), \(Z_\gamma\) converges in distribution to a
\(\operatorname{Poisson}(1)\) random variable and \(D_\gamma\to 1/(2\exp(1))\).
\end{proposition}

At unit concentration the exact gap is
\begin{equation}
 D_1=\frac{e^{-\gamma_{\mathrm E}}}{4}\simeq0.1404,
 \label{ex:fixed-unit}
\end{equation}
where \(\gamma_{\mathrm E}\) is Euler's constant.

\paragraph{Confidence-sensitive reserves.}
\label{ex:fixed-confidence}
Suppose upstream debtors pay \(q\in(0,1)\) and dealers in \(V\) respond to
missed receipts with \(\nu>0\). A dealer owed \(\bar z\) receives
\(q\bar z\) and misses \((1-q)\bar z\). It repays \(\min\{1,a+b\bar z\}\),
where \(a=1-h_V\in[0,1)\) and \(b=q-\nu(1-q)>0\). The continuum uses
\(\bar z=1\), while the finite limit averages over \(\bar z=Z_\gamma\).
The cap gives a strictly positive gap,
including when \(a+b>1\) and continuum dealers have a strict full-payment
margin. Appendix~\ref{app:fixed-gamma} derives the extension and
reports numerical values for \(D_\gamma\).

\section{Network comparisons and stress testing}
\label{ex:blocks}

\subsection{Same balance-sheet totals, different partial freezes}
\label{ex:matched-marginals}

Consider two liquidity-sensitive dealers \(A,B\) and two cash-rich clearing
banks \(C,D\), each with \(\bar p_i=\bar z_i=1\). The dealers have
\((e_i,h_i,\nu_i)=(2/5,1/10,1/2)\), and the clearing banks have
\((e_i,h_i,\nu_i)=(1,0,0)\). For example, the first pair of reserve
parameters is obtained with \(c_i=1\), \(\omega_i=1/10\) and \(\psi_i=1/2\),
and the second pair with \(c_i=0\). In each network below, an arrow of
unit weight points from a debtor to its creditor. Both liability matrices
belong to \(\mathcal L(\bar p,\bar z)\) in \eqref{samp:feasible-margins}.

By \eqref{mod:best}, banks \(C,D\) always pay one. A bank in \(A,B\) that
receives repayment \(u\) from its debtor chooses
\(f(u)=\clip(-1/5+3u/2)\). With separate cycles \(A\leftrightarrow B\) and
\(C\leftrightarrow D\), the two fragile banks must pay the same amount.
Indeed, \(s_A<s_B\) would imply \(s_B=f(s_A)\leq f(s_B)=s_A\), and the
reverse strict inequality is excluded in the same way. The fixed points
of \(f\) are \(0,2/5,1\). In the alternating cycle
\(A\to C\to B\to D\to A\), each fragile bank receives payment from one
of \(C,D\), so it chooses \(f(1)=1\). Table~\ref{ex:matched-marginals-table}
reports all the resulting strategic equilibria.

\begin{table}[htbp]
\centering
\caption{Strategic repayments under two same-margins reconstructions}
\label{ex:matched-marginals-table}
\begin{tabular}{@{}lll@{}}
\toprule
Liability links & Equilibria in order \(A,B,C,D\) & Least aggregate repayment\\
\midrule
\(A\leftrightarrow B,\ C\leftrightarrow D\)
  & \((0,0,1,1),\ (2/5,2/5,1,1),\ \mathbf 1\) & \(1/2\)\\
\(A\to C\to B\to D\to A\)
  & \(\mathbf 1\) & \(1\)\\
\bottomrule
\end{tabular}
\end{table}

Mechanical clearing uniquely yields full early payment throughout this
feasible set. At a mechanical fixed point, the unpaid fraction
\(d_i=1-s_i\) satisfies \(d_i=[\sum_jL_{ji}d_j-e_i]_+\) by
\eqref{mod:mechanical} and the unit claim totals. If
\(D_+=\{i:d_i>0\}\) were nonempty, the unit row sums would give
\(\sum_{i\in D_+}d_i\leq\sum_j d_j-\sum_{i\in D_+}e_i\), which is
impossible because each \(e_i>0\).

The strategic corner tests also agree across all reconstructions. Full
payment satisfies \eqref{samp:full-margin}, and \(C,D\) violate
\eqref{samp:zero-margin}. Since \(C,D\) pay one in every strategic
equilibrium, the least aggregate repayment rate lies between \(1/2\) and
\(1\) by \eqref{mod:rr}. Both bounds are attained by the networks in
Table~\ref{ex:matched-marginals-table}.

\subsection{Sharpening and the freeze threshold}
\label{ex:sharpening}

The remaining comparisons take a population kernel as given. Contractual
claims may vary across networks sampled under the canonical allocation rule.

Three debtor blocks and one pure-creditor block each have mass \(1/4\).
Debtors have \(\bar p=1\) and allocate shares \(13/25\), \(3/25\) and \(9/25\)
to the successor in the directed cycle, the predecessor and the pure
creditors, respectively. The kernel equals four times the relevant share
on each block rectangle. Sparsemax acts on the vector of block shares
divided by \(\tau\). With equal block masses, cyclic routing and unit
debts, the total share \(r_\tau\) that each debtor allocates to other
debtors also equals each debtor's contractual claims at the target
\(P_\tau\). It satisfies
\begin{equation}
r_\tau =
\begin{cases}
\frac{2}{3}-\frac{2}{75\,\tau}, & 16/25\leq\tau\leq1,\\[2pt]
\frac12+\frac{2}{25\,\tau}, & 4/25\leq\tau<16/25,\\[2pt]
1, & 0<\tau<4/25 .
\end{cases}
\label{ex:active-share}
\end{equation}
The predecessor link vanishes at \(\tau=16/25\), and the pure-creditor
share vanishes at \(\tau=4/25\). With common debtor characteristics,
the feedback radius is \((1+\nu)r_\tau\), since the positive block
operator has this constant row sum.

\begin{figure}[tbp]
 \centering
 \includegraphics[width=0.84\linewidth]{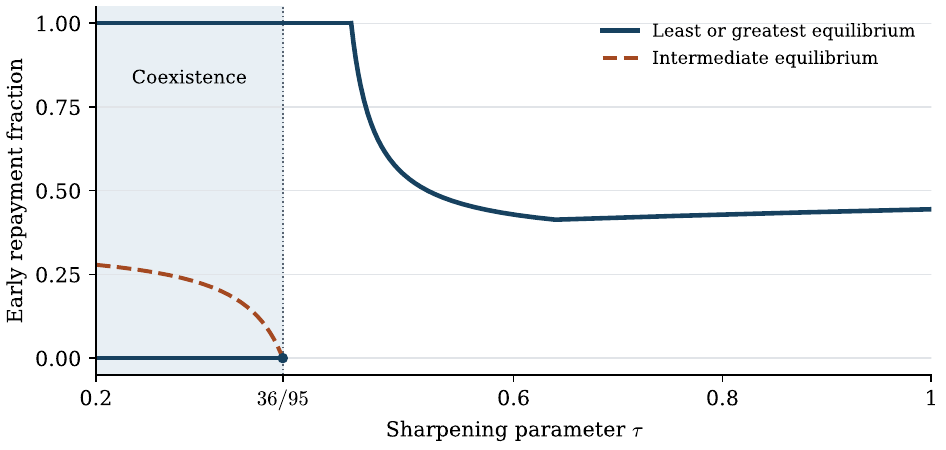}
 \caption{Strategic repayment branches in the four-block economy.
 Shading marks coexistence of a freeze and full payment.}
 \label{ex:branches}
\end{figure}

Set \((e,h,\nu)=(13/25,1/5,9/20)\). At \(\tau=1\), contraction gives
unique repayment \(4/9\). For \(\tau\leq16/25\) only the successor link
remains among debtors, and the block response to its predecessor's
repayment \(u\) is
\[
 T_\tau(u)=\clip
 \left(\frac{8}{25}-\frac{9r_\tau}{20}+\frac{29r_\tau}{20}u\right).
\]
Every fixed point of \(T_\tau^3\) is a fixed point of \(T_\tau\) by
monotonicity, so equilibrium repayment is symmetric. A freeze exists
for \(r_\tau\geq32/45\), and full strategic repayment exists for
\(r_\tau\geq17/25\). The two corners coexist at and below
\(\tau=36/95\), with an additional interior branch strictly below that
cutoff. Above the cutoff, contraction gives uniqueness when
\(r_\tau<20/29\), and the remaining scalar maps have positive intercept
and slope at least one, giving unique full payment. For example, the
three strategic repayments at \(\tau=7/20\) are \(0\), \(11/79\) and \(1\).
Mechanical clearing stays unique and full in Figure~\ref{ex:branches},
where \(r_\tau<1\) and \(e+r_\tau>1\).

With \((e,h,\nu)=(9/20,1/4,3/20)\), the settings
\(\tau=16/25\) and \(\tau=2/5\) both remove the predecessor link, but
change continuum repayment by \(-1.62\) and \(+9.32\) percentage points relative
to \(\tau=1\). Sparsemax changes the weights on the retained links and
therefore changes incoming claims at fixed debtor totals.
Appendix~\ref{ex:targets} separates changes in the target from sampling
errors using paired simulations.

\subsection{Finite-network survival of the repayment branches}
\label{ex:coexistence-simulation}

For a strict-margin calibration, set \(\tau=1/5\) and
\((e,h,\nu)=(18/25,1/5,23/20)\). Nominal inflows are \(9/10\), the
freeze and full-payment margins are \(103/200\) and \(21/50\), and the
mechanical radius is \(9/10\). The intermediate repayment \(103/187\) satisfies
Theorem~\ref{pers:noncontractive}: the nonzero feedback eigenvalues
are \(387/200\) times the cube roots of unity, so the spectrum excludes
one even though the radius exceeds one.

We sample 48 networks per size and concentration regime according to
\eqref{samp:sampler}. We test the exact corner conditions and recover
the least and greatest equilibria by monotone iteration where needed.
Mechanical clearing is unique in every draw because active banks
have strictly positive opening cash. Appendix~\ref{app:coexistence}
gives the argument and replication details. All 160 networks tested for the
intermediate branch yielded profiles strictly interior on the active
debtor coordinates, with maximum clipped residual at most \(10^{-9}\).

\begin{figure}[!tb]
 \centering
 \includegraphics[width=0.96\linewidth]{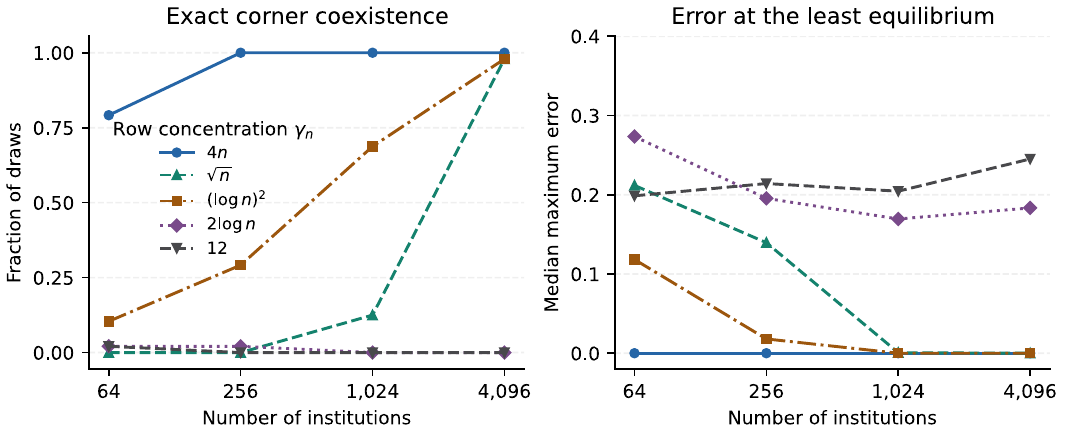}
 \caption{Exact coexistence frequency and median maximum error at the least
 strategic equilibrium across 48 draws per point.}
 \label{ex:coexistence-figure}
\end{figure}

\begin{table}[!tb]
 \centering
 \caption{Corner equilibria and exposure concentration at \(n=4{,}096\)}
 \label{ex:coexistence-table}
 \small
 \begin{tabular}{lrrrrrr}
 \toprule
 \(\gamma_n\) & Both & Mech.\ full & Least error & Greatest error & Effective & Top share \\
 \midrule
 \(4n\)         &48&48&0&0&1159.7&\(0.2\%\)\\
 \(\sqrt n\)    &47&48&0&0&63.3&\(5.4\%\)\\
 \((\log n)^2\) &47&48&0&0&68.0&\(5.0\%\)\\
 \(2\log n\)    &0&34&0.183&0.187&18.7&\(14.0\%\)\\
 \(12\)         &0&8&0.245&0.241&14.1&\(17.4\%\)\\
 \bottomrule
 \end{tabular}
 \smallskip

 \parbox{0.95\linewidth}{\small Counts are out of 48. Errors are medians of
 the maximum deviations from the respective continuum corners. Exposure
 columns give medians across draws of the active-debtor averages of
 \(1/\sum_j\Pi_{ij}^2\) and \(\max_j\Pi_{ij}\).}
\end{table}

With \(\gamma_n=4n\), both corners survive in 38 of 48 draws at \(n=64\)
and 48 of 48 at every larger size. At \(n=4{,}096\), neither
\(\gamma_n=2\log n\) nor \(\gamma_n=12\) yields both exact corners in any draw
(Figure~\ref{ex:coexistence-figure} and
Table~\ref{ex:coexistence-table}). With \(\gamma_n=12\), the median aggregate
extrema are \(0.0016\) and \(0.9977\), with median maximum bank-level
deviations above \(0.24\) at both ends. In this calibration, low receipts
rule out both exact corners with probability tending to one at any fixed
\(\gamma>0\) (Appendix~\ref{app:coexistence}). For \(\gamma_n=2\log n\),
Theorem~\ref{pers:noncontractive} gives exact finite equilibria approaching
each continuum corner in \(L^2\).

\paragraph{Reading the stress test.}
Across reconstructions with fixed balance-sheet totals and marks, report
the least and greatest strategic repayment rates and which institutions
withhold, together with mechanical clearing on each matrix. For draws
conditional on a fitted kernel, report the deviations in contractual
claims alongside the repayment statistics. Aggregate repayment can be
close to either continuum corner even when individual banks deviate
substantially from it.

\subsection{A core--periphery stress test}
\label{ex:core-periphery}

The Bank of England's system-wide exploratory scenario examines how the
responses of banks and non-banks amplify shocks \citep{boeswes2024}.
We isolate a payment-timing channel in a synthetic network with a dealer
core, a periphery of smaller banks, and downstream cash creditors. The
three groups have masses \(10\%\), \(70\%\), and \(20\%\). Each core bank owes
four times a periphery bank's obligations and has a confidence-sensitive
reserve target. Periphery banks can pay in full from opening cash after
retaining reserves. The cash creditors have no outgoing obligations.

Use the block-affinity specification of
Section~\ref{subsec:estimation-bridge}, with unequal sender and receiver
affinities. At \(\tau=1\) the graphon has a unique strategic equilibrium
with \(94.97\%\) aggregate early repayment. At \(\tau=1/5\), the share of a
core bank's liabilities owed to the core rises from \(30\%\) to \(86.67\%\).
The resulting equilibrium set includes full payment and a partial freeze
in which the core pays zero and the periphery pays in full. Mechanical
clearing remains uniquely full. Appendix~\ref{app:core-periphery}
gives the affinities, bank parameters and equilibrium calculations.

Construct a maximum-entropy kernel at the liability and claim margins of the
sharpened target \(P_{1/5}\). It allocates each debtor's obligations in
proportion to creditor claims, yielding unique full payment under both
strategic and mechanical clearing. The share of a core bank's claims
owed by other core banks rises from \(4/11\) under maximum entropy to
\(208/271\) under \(P_{1/5}\). The least strategic repayment rate falls
from one to \(7/11\). These are the sharp
bounds at the common margins because the periphery pays in full in every
strategic equilibrium.

We draw \(120\) finite networks with \(600\) institutions at the stated
block proportions. At \((\tau,\gamma)=(1/5,16)\), the median active-row mean
effective counterparty count \(1/\sum_j\Pi_{ij}^2\) is \(17.18\). Periphery
banks pay in full at the least equilibrium in every draw, and the median
of the within-core repayment rate is \(0.23\%\). The 5th and 95th percentiles of least aggregate
repayment across draws are \(63.64\%\) and \(63.93\%\). Mechanical clearing
pays in full on each of the same matrices. Table~\ref{ex:core-periphery-table}
reports the equilibrium extrema.

\begin{table}[!htbp]
\centering
\caption{Early repayment in the synthetic core--periphery economy (percent)}
\label{ex:core-periphery-table}
\small
\setlength{\tabcolsep}{4pt}
\begin{tabular}{@{}lrrr@{}}
\toprule
Model & \shortstack[r]{Core,\\least} & \shortstack[r]{Aggregate,\\least} & \shortstack[r]{Aggregate,\\greatest} \\
\midrule
Unsharpened graphon \(P_1\), strategic & 86.17 & 94.97 & 94.97 \\
\addlinespace
\multicolumn{4}{@{}l}{\textit{Graphons at the same sharpened margins}} \\
Sharpened graphon \(P_{1/5}\), mechanical & 100.00 & 100.00 & 100.00 \\
Maximum entropy, strategic & 100.00 & 100.00 & 100.00 \\
Sharpened graphon \(P_{1/5}\), strategic & 0.00 & 63.64 & 100.00 \\
\addlinespace
\multicolumn{4}{@{}l}{\textit{Numerical equilibria for draws around the sharpened kernel}} \\
Sampled \((\tau,\gamma)=(1/5,16)\), strategic median & 0.23 & 63.72 & 100.00 \\
\bottomrule
\end{tabular}
\end{table}

The sampler preserves debtor obligations. Relative to the sharpened
graphon, the median sum across banks of absolute differences in creditor
claims is \(19.86\%\) of total liabilities. The comparison at common
margins isolates bilateral wiring. The finite draws show how the stress
scenario behaves under allocation dispersion at fixed debtor totals.

\section{Conclusion}
Given opening liquidity and reserve parameters, balance-sheet totals
determine whether zero and full early payment are equilibria. To assess
partial freezes, stress tests must also vary who owes whom across feasible
reconstructions.

With growing concentration, strict margins and an invertible interior
equilibrium equation give exact finite counterparts to graphon equilibria,
including some noncontractive branches. At fixed concentration, dispersed receipts and the payment cap
can produce a persistent shortfall in early repayment even when all
principal settles. Stress tests should track both aggregate repayment and
which banks withhold payment.

\bibliographystyle{plainnat}
\begingroup
\fontsize{9}{10}\selectfont
\setlength{\bibsep}{0pt}
\setlength{\bibhang}{10pt}
\bibliography{references,extra_references}
\endgroup

\clearpage
\appendix
\WorkingPaperAppendixHeadings
\counterwithin{equation}{section}
\counterwithin{figure}{section}
\counterwithin{table}{section}
\renewcommand{\theHequation}{appendix.\thesection.\arabic{equation}}
\renewcommand{\theHfigure}{appendix.\thesection.\arabic{figure}}
\renewcommand{\theHtable}{appendix.\thesection.\arabic{table}}

\section{Proofs}
\label{app:proofs}
\subsection{Proofs for the repayment game}
\label{app:model-proofs}

\begin{proof}[Proof of Proposition~\ref{mod:bestresponses}]
For a given type and profile, use the type-indexed counterparts of \eqref{mod:resources} and \eqref{mod:netpayoff}. Then optimising over early payments is equivalent to optimising over \(R\in[(a-\bar p)_+,a]\). Up to an additive constant in \(R\), the objective in \eqref{mod:netpayoff} is \(\chi\int_0^R\overline G_x(t/\kappa)\,dt-\delta R\). Its derivative \(\chi\overline G_x(R/\kappa)-\delta\) is strictly decreasing. The unique optimal reserve is the target \(c\kappa\) clipped to the feasible reserve interval. Rewriting the resulting payment gives \eqref{mod:best}, using \([[e+Ks]_+-c\kappa]_+=[e+Ks-c\kappa]_+\). Positivity of \(M\) and the monotonicity and unit Lipschitz constant of the clip give \eqref{mod:domination} and the order result.

For the embedded economy, the conditional inflow in \eqref{mod:embedding} is constant on each cell and depends on other cells because \(L_{ii}=0\). Within each cell \eqref{mod:best} equals the finite best response to the vector \((\langle s\rangle_j)_{j=1}^n\) of cell averages. Every fixed point is therefore a step profile, and a step profile is a fixed point precisely when its vector is a finite equilibrium.
\end{proof}

\begin{proof}[Proof of Theorem~\ref{mod:lattice}]
The almost-everywhere order makes \(\mathcal S_A\) a complete lattice, using essential suprema and infima. The order-preserving map \(B\) has a nonempty complete lattice of fixed points by Tarski's fixed-point theorem \citep{tarski1955}. Iteration from the top is decreasing and bounded. The dominated convergence theorem gives convergence in \(L^2\) to the almost-everywhere limit. For almost every receiver, the integrable function \(P(y,x)\bar p(y)\) dominates the integrand in \eqref{mod:inflow}. Passing to the limit under the integral and using continuity of the clip shows that the limit is a fixed point. Every fixed point is bounded by each iterate from the top, which identifies the limit as greatest. Iteration from zero is increasing and yields the least fixed point by the same reasoning. The map \(\Phi\) has the same structure. Since desired reserves are nonnegative, \(B\leq\Phi\) pointwise. Comparing their iterates from each common endpoint gives \eqref{mod:comparison}. Under the stated discipline threshold both \(h\) and \(\nu\) are zero.
\end{proof}

\begin{proof}[Proof of Proposition~\ref{mod:spectral}]
For two initial profiles, \eqref{mod:domination} gives \(|B^k(s)-B^k(t)|\leq M^k|s-t|\) by induction. When \(\rho(M)<1\), the spectral radius formula implies \(\|M^k\|_{2\to2}\to0\). Taking \(t\) to be an equilibrium proves convergence and uniqueness.

The multiplication operator \(D_{\bar p}:f\mapsto\bar p f\) is bounded and invertible on \(L^2(A)\). The adjoint of \(T_{AA}\) satisfies \(T_{AA}^*=D_{\bar p}M_0D_{\bar p}^{-1}\). To bound the radius of this compact operator, let \(\lambda\) be any nonzero eigenvalue of \(T_{AA}^*\) with eigenfunction \(f\). Positivity gives \(|\lambda||f|\leq T_{AA}^*|f|\). Integrating and using the rows of \(P\) shows that
\[
 |\lambda|\int_A |f|
 \leq \int_A (1-\ell_{\mathrm{out}})|f|
 \leq (1-\operatorname*{ess\,inf}_{A}\ell_{\mathrm{out}})\int_A |f|.
\]
Every nonzero spectral value of this compact operator is an eigenvalue.
Since \(\rho(T_{AA}^*)=\rho(T_{AA})\) and similarity preserves the spectrum,
this gives \eqref{mod:leakage}. The comparison \(0\leq M\leq (1+\|\nu\|_\infty)M_0\) implies the same order for the corresponding powers and bounds their operator norms. Taking \(k\)-th roots gives the sufficient condition for strategic uniqueness. Applying the first part with \(M_0\) yields mechanical uniqueness under uniform leakage.
\end{proof}

\paragraph{The mechanical limit.}
Holding the kernel and other primitives fixed, if \(\delta_m\) increases and
\(c_m\downarrow0\) almost everywhere, the response
maps increase pointwise and remain bounded above by \(\Phi\). Their extremal
fixed points increase by the endpoint iterations. Dominated convergence in
\(K\), together with \(h_m=\omega c_m\to0\) and \(\nu_m=\psi c_m\to0\), makes
each limit a fixed point of \(\Phi\). If it is unique, bounded convergence
gives \(L^2\) convergence of both strategic extrema.

\subsection{Stability under changes in the liability kernel}
\label{stab:section}

The following stability result compares kernels after aligning the types and their
balance-sheet and behavioural characteristics. For an integrable signed kernel
$D$, use the directed cut norm
\[
 \|D\|_{\square}=
 \sup_{S,T\subseteq[0,1]}
 \left|\int_{S\times T}D(x,y)\,\mathrm dx\,\mathrm dy\right|,
\]
where the supremum runs over measurable sets. The sender and receiver sets may
differ.

\begin{proposition}[Stability under aligned cut-norm perturbations]
\label{stab:cut}
Fix the type characteristics. Let $P_m,P$ be bounded row-normalised liability
kernels on a common type space, with $\|P_m-P\|_{\square}\to0$. Every sequence of
equilibria under $P_m$ has a subsequence converging in $L^1$ to an equilibrium
under $P$. If the equilibrium under $P$ is unique, any sequence of such equilibria
converges to it in $L^1$, and the corresponding aggregate repayment rates converge.
\end{proposition}

\begin{proof}
Let $B_m$ and $B$ be the best-response maps for $P_m$ and $P$, respectively.
Let $K_D$ be the inflow operator \eqref{mod:inflow} with signed kernel $D=P_m-P$.
For $0\le f\le 1$, use the layer-cake decomposition of $\bar p f$ and the positive and
negative parts of the sign of $K_D f$ to obtain
$\|K_D f\|_1\le 2\|\bar p\|_\infty\|D\|_{\square}$. Applying the bound also to
$\mathbf 1_A$ controls nominal inflows. The lower bound on active $\bar p$,
boundedness of $\nu$ and Lipschitz clipping imply
$\sup_{s\in\mathcal S_A}\|B_m(s)-B(s)\|_1\to0$.
The limiting map $B$ has relatively compact image in $L^1$: its linear part has
a square-integrable kernel, so it is compact on $L^2$, and clipping preserves
relative compactness there and hence in $L^1$. For a sequence of fixed points
$s_m$, the bound $\|s_m-B(s_m)\|_1\to0$ gives an $L^1$-convergent subsequence.
The limit lies in $\mathcal S_A$ and is fixed by $B$, since the bounded limiting
kernel gives $L^1$ continuity. Uniqueness identifies every subsequential limit.
Integration against the bounded obligation function is continuous in $L^1$.
\end{proof}

The result concerns aligned marked economies. A relabelling of the kernel used
to compare two economies must also be applied to their type characteristics.

\subsection{Sampling bounds and equilibrium approximation}
\label{samp:proof-appendix}

\noindent\textit{Receipt concentration.}\enspace
Put \(b_n=\sqrt{\log n/n}\) and \(d_n=\sqrt{\log n/\gamma_n}\), with
\(d_n=0\) for deterministic allocations. In the bounds below, a maximum of
nonnegative quantities over an empty index set is zero.

\begin{lemma}[Uniform approximation of receipts]
\label{samp:inflows}
For a fixed finite collection \(\mathcal F\) of \(m\) Borel functions taking
values in \([0,1]\), fix \(\ell>0\) and define
\[
 \begin{gathered}
 H=P_{\max}/\tau,\qquad C_w=2H,\qquad
 \epsilon=H(1+2P_{\max})\sqrt{\ell/[2(n-1)]},\\
 a_\gamma=\sqrt{2\ell/\gamma}+2\ell/\gamma,\qquad
 d_\gamma=\sqrt{2C_w\ell/\gamma}+2\ell/\gamma .
 \end{gathered}
\]
At any \(\gamma\in(0,\infty]\) with \(\epsilon\le1/2\) and \(a_\gamma<1\),
\begin{equation}
\begin{gathered}
 \max_{f\in\mathcal F}\max_{1\le i\le n}|(K_nR_nf)_i-(K_\tau f)(x_i)|\le E,\\
 E=p_{\max}\left((1+2C_w)\epsilon+
                      \frac{d_\gamma+C_wa_\gamma}{1-a_\gamma}\right),
\end{gathered}
 \label{samp:receipt-constant}
\end{equation}
except on an event of probability at most \((6+4m)n e^{-\ell}\). At
\(\gamma=\infty\) use \(a_\gamma=d_\gamma=0\).

If \(\gamma_n/\log n\to\infty\), for any fixed finite collection
\(\mathcal F\) of bounded Borel functions and any \(q>1\) there are
finite deterministic \(C_q,D_q\) such that for all sufficiently large \(n\),
\begin{equation}
 \Pr\left\{\max_{f\in\mathcal F}\max_{1\le i\le n}
       |(K_nR_nf)_i-(K_\tau f)(x_i)|>C_qr_n\right\}\le D_qn^{-q}.
 \label{samp:inflow-bound}
\end{equation}
The bound holds eventually almost surely under any coupling across network sizes.
\end{lemma}

\begin{proof}[Proof of Lemma~\ref{samp:inflows}]
Put \(N=n-1\) and \(u=\sqrt{\ell/(2N)}\). On each active row let
\(D_{n,k}=N^{-1}\sum_{j\ne k}P(x_k,x_j)\). Conditional Hoeffding bounds yield
\(|D_{n,k}-1|\le P_{\max}u\) and an error of at most \(Hu\) in the empirical
mean of \([P(x_k,x_j)/\tau-\theta_\tau(x_k)]_+\) from its population value one.
The joint failure probability is at most \(4ne^{-\ell}\).
On this event \(D_{n,k}\ge1/2\). Dividing the scores by \(D_{n,k}\) changes the
excess mean by at most \(2HP_{\max}u\), leaving it within \(\epsilon\) of one
at \(\theta_\tau(x_k)\). Between this threshold and the empirical root at
level one, the excess mean is at least \(1-\epsilon\), and the scores are
bounded by \(C_w\). The fraction of scores above the threshold is therefore
at least \((1-\epsilon)/C_w\) throughout this interval. Integrating that
fraction between the thresholds bounds their difference by \(2C_w\epsilon\).
Thus
\begin{equation}
 \max_{k\in A_n,\,j\ne k}|Nw_{kj}-P_\tau(x_k,x_j)|
 \le 2HP_{\max}u+2C_w\epsilon,\qquad
 \max_i\sum_{k\ne i} w_{ki}\le C_w .
 \label{samp:mean-weights}
\end{equation}
For each \(f\), bound the error of the ideal empirical column average of
\(P_\tau(x_k,x_i)\bar p(x_k)f(x_k)\) from \((K_\tau f)(x_i)\) by
\(p_{\max}Hu\) using Hoeffding conditional on \(x_i\). The failure probability
over columns and functions is at most \(2mn e^{-\ell}\). Using
\eqref{samp:mean-weights}, the total error of the mean-weight receipts is
at most \(p_{\max}(1+2C_w)\epsilon\).

At \(\gamma=\infty\) this already proves the claimed error bound.
For finite \(\gamma\), draw conditionally independent
\(G_{ki}\sim\operatorname{Gamma}(\gamma w_{ki},1)\) on the positive support
and set \(G_{ki}=0\) elsewhere. On active rows let \(G_{k+}=\sum_iG_{ki}\), so
\(G_{k+}\sim\operatorname{Gamma}(\gamma,1)\) and \(\Pi_{ki}=G_{ki}/G_{k+}\).
A gamma variate of shape \(b\) has centred log moment generating function
bounded by \(b\lambda^2/[2(1-|\lambda|)]\) for \(|\lambda|<1\).
Consequently \(|G_{k+}/\gamma-1|\le a_\gamma\) simultaneously on active rows
with conditional failure probability at most \(2ne^{-\ell}\). The same
bound and \eqref{samp:mean-weights} give
\(|z^{\mathrm{raw}}_{i,f}-\mathbb E[z^{\mathrm{raw}}_{i,f}\mid\boldsymbol{x}]|\le p_{\max}d_\gamma\) simultaneously
for \(z^{\mathrm{raw}}_{i,f}=\gamma^{-1}\sum_k\bar p(x_k)f(x_k)G_{ki}\), with conditional
failure probability at most \(2mn e^{-\ell}\). The variance factor and
scale in the latter tail bound are \(p_{\max}^2C_w/\gamma\) and
\(p_{\max}/\gamma\). Because \(z^{\mathrm{raw}}_{i,f}\ge0\), restoring row normalisation
changes it by at most \(a_\gamma z^{\mathrm{raw}}_{i,f}/(1-a_\gamma)\), and its conditional
mean is at most \(p_{\max}C_w\). Summing the error bounds and failure
probabilities gives \eqref{samp:receipt-constant}.

For bounded signed \(f\), apply this bound to its positive and negative parts
divided by \(\max\{1,\|f\|_\infty\}\), then add the two rescaled errors.
Take \(\ell=(q+1)\log n\). The growth condition gives
\(\epsilon=O(b_n)\) and \(a_\gamma,d_\gamma=O(d_n)\), so
\eqref{samp:inflow-bound} follows. Borel--Cantelli gives the almost-sure
claim.
\end{proof}

\subsubsection*{Alignment of sampled types}

For a continuum equilibrium profile \(s\) and sampled vector \(u\),
\begin{equation}
 \|\widetilde u_n-s\|_{L^2}
 \le \max_{1\le i\le n}|u_i-s(x_i)| + \|s\circ T_n-s\|_{L^2}.
 \label{samp:profile-bound}
\end{equation}
The first term compares actions at sampled types. The second records the error from representing a profile on equal cells.

\begin{lemma}[Representation of continuum profiles]
\label{samp:step}
For any fixed bounded Borel profile \(f\), \(\|f\circ T_n-f\|_{L^2}\to0\) almost surely. If \(f\) is Lipschitz on each interval of a finite partition of \([0,1]\), this error is \(O((\log n/n)^{1/4})\) almost surely. If \(f\) is globally Lipschitz, it is \(O((\log n/n)^{1/2})\) almost surely.
\end{lemma}

\begin{proof}[Proof of Lemma~\ref{samp:step}]
Let \(F_n^{\mathrm{emp}}\) be the empirical distribution function of the sampled types. At each point of the grid \(\{j/n\}_{j=0}^n\), Hoeffding's inequality bounds the deviations of \(F_n^{\mathrm{emp}}\) from the uniform distribution function. A union bound and monotonicity of \(F_n^{\mathrm{emp}}\) between grid points give
\(\|F_n^{\mathrm{emp}}-\mathrm{id}\|_\infty=O((\log n/n)^{1/2})\) almost surely by Borel--Cantelli. Since there are almost surely no ties, \(F_n^{\mathrm{emp}}(x_{(i)})=i/n\). Thus \(\delta_n=\|T_n-\mathrm{id}\|_\infty\) has the same bound. The claim follows for continuous \(f\) by uniform continuity. For a bounded measurable \(f\), take a sequence of continuous \(g_k\) converging to \(f\) in \(L^2\). For each \(k\),
\[
 \|(f-g_k)\circ T_n\|_{L^2}^2
   =\frac1n\sum_{i=1}^n|f(x_i)-g_k(x_i)|^2
   \longrightarrow \|f-g_k\|_{L^2}^2
   \quad\text{almost surely}.
\]
This follows from Hoeffding's inequality and Borel--Cantelli for any coupling across \(n\). On the common probability-one event, the triangle inequality gives \(\limsup_n\|f\circ T_n-f\|_{L^2}\le 2\|f-g_k\|_{L^2}\) for every \(k\). For the piecewise Lipschitz rate, outside the \(\delta_n\)-neighbourhood of the finitely many partition points, \(x\) and \(T_n(x)\) lie in the same piece and the difference is \(O(\delta_n)\). Those neighbourhoods have total length \(O(\delta_n)\), and the profile is bounded there, giving \(O(\delta_n^{1/2})\) in \(L^2\). Global Lipschitz continuity gives \(O(\delta_n)\) everywhere.
\end{proof}

\subsubsection*{Equilibrium sets and nondegenerate branches}

\begin{proof}[Proof of Theorem~\ref{samp:ne-limit}]
We first show that the realised step liability kernel \(P_n\), with entries
\(n\Pi_{ij}\) ordered by the latent types, converges to \(P_\tau\) in cut norm
almost surely. The corresponding mean-weight kernel \(P_n^{\mathrm{mean}}\), with entries
\(nw_{ij}\) in the same order, converges to \(P_\tau\) in \(L^1\). Indeed, the mean-weight bound
\eqref{samp:mean-weights} uses only the types and holds eventually almost surely
by the summable Hoeffding bounds. It gives uniform approximation by the
off-diagonal entries \(P_\tau(x_i,x_j)\), since \(n/(n-1)\to1\).
The diagonal cells have total area \(1/n\). To see that
\(P_\tau(T_n(x),T_n(y))\to P_\tau(x,y)\) in \(L^1\), first use uniform continuity
for a continuous kernel. For a bounded measurable kernel, approximate it in
\(L^1\) by continuous kernels. For any fixed bounded Borel error function
\(H\) on \([0,1]^2\), the empirical average \(n^{-2}\sum_{i,j}H(x_i,x_j)\)
converges almost surely to \(\int_{I^2}H(x,y)\,\mathrm dx\,\mathrm dy\). Its expectation differs from the
integral by \(O(n^{-1})\), due to the diagonal. Changing a single type
alters at most \(2n\) bounded terms, so the bounded-differences inequality
gives a summable probability of any fixed deviation. Taking \(H\) to be
the absolute approximation error and using a countable sequence of continuous
approximants proves the \(L^1\) claim, as in Lemma~\ref{samp:step}.

At sizes with \(\gamma_n=\infty\), \(P_n=P_n^{\mathrm{mean}}\) exactly. At sizes with
finite \(\gamma_n\), use the independent conditional gamma variables \(G_{ij}\)
from Lemma~\ref{samp:inflows}, with inactive rows set to zero. For any coefficients \(a_{ij}\in\{-1,0,1\}\), the conditional
log moment generating function of the centred sum
\(\sum_{i,j}a_{ij}(G_{ij}-\gamma_n w_{ij})\) is bounded by
\(n\gamma_n\lambda^2/(2(1-|\lambda|))\) for \(|\lambda|<1\), because the total
shape is at most \(n\gamma_n\). The conditional probability that the sum
divided by \(n\gamma_n\) exceeds \(t>0\) in absolute value is at most
\(2\exp[-n\gamma_n t^2/(2(1+t))]\).
For a step kernel the cut norm maximisation reduces to subsets of cell indices,
since the integrals are bilinear in the fractions of each cell included.
Thus there are at most \(4^n\) cuts. Taking
\(a_{ij}=\mathbf1_{\{i\in S,\,j\in T\}}\) and \(t=C/\sqrt{\gamma_n}\) gives an
eventually summable failure probability for the maximum over cuts, for a large
enough \(C\). The cut error from normalising the gamma rows is bounded by
\((n\gamma_n)^{-1}\sum_{i\in A_n}|G_{i+}-\gamma_n|\), where
\(G_{i+}=\sum_jG_{ij}\). Express the sum of absolute values as a maximum over
at most \(2^n\) row signs and apply the same bound with \(a_{ij}\) equal to
the sign on row \(i\). It follows that
\(\|P_n-P_n^{\mathrm{mean}}\|_\square=O(\gamma_n^{-1/2})\) almost surely, with zero allocation
error at deterministic sizes, and so
\(\|P_n-P_\tau\|_\square\to0\).

The sampled balance-sheet and reserve characteristics also converge as step
functions in \(L^1\) by Lemma~\ref{samp:step}. This applies to \(\bar p,e,h,\nu\)
and the bounded reciprocal of \(\bar p\) on \(A\), extended by zero off \(A\).
Set \(h\) and \(\nu\) to zero off \(A\) for the step comparison. Define the continuum and step best responses on all \([0,1]\)-valued
profiles by using the corresponding inverse liability in the clipped index,
which gives zero off the respective active set. To compare their inflows,
first replace \(P_n\) by \(P_\tau\) at the sampled obligation schedule and then
replace that schedule by \(\bar p\). The uniform \(L^1\) error for \(0\le f\le1\)
is bounded by
\(2p_{\max}\|P_n-P_\tau\|_\square
 +\|P_\tau\|_\infty\|\bar p\circ T_n-\bar p\|_1\)
by the cut estimate of Proposition~\ref{stab:cut}. The same bound applies
to nominal inflows with \(f=1\). Limiting inflows are uniformly bounded.
Together with convergence of the other characteristics and Lipschitz clipping,
this gives uniform \(L^1\) convergence of the step best responses to the
continuum response on that strategy set.

The limiting response has relatively compact image in \(L^1\), as in
Proposition~\ref{stab:cut}. On the probability-one event just constructed,
every sequence of step fixed points therefore has an \(L^1\)-convergent
subsequence whose limit is a continuum fixed point. Both step and continuum
profiles take values in \([0,1]\), so \(L^1\) convergence implies \(L^2\)
convergence. If \eqref{samp:ne-set} failed, a sequence of step fixed points
would stay a positive \(L^2\) distance from \(\mathcal E_\tau\), contradicting
the subsequence conclusion. A singleton \(\mathcal E_\tau\) gives
\eqref{samp:ne-unique}.
\end{proof}

\begin{proof}[Proof of Theorem~\ref{pers:noncontractive}]
Extend the clipped affine formulas for \(B_\tau\) and the sorted step response
\(B_n^{\rm step}\) to all of \(L^2(I)\), setting responses to zero off their
respective active sets. Let \(C\) clip input profiles to \([-1,2]\), and define
\(\widehat B=B_\tau\circ C\) and \(\widehat B_n=B_n^{\mathrm{step}}\circ C\).
These maps are continuous. Their outputs are in \([0,1]\), so their
fixed points are unchanged by input clipping. The limit map is compact
because \(M_\tau\) is Hilbert--Schmidt and the input and output clips are
Lipschitz. The finite map has bounded image in the \(n\)-dimensional space
of step functions.

The proof of Theorem~\ref{samp:ne-limit} gives uniform \(L^1\) convergence
of the response maps for inputs valued in \([-1,2]\) as well. The cut norm
estimate applies separately to the positive and negative parts of each
bounded input, and the errors from sampling the type characteristics have
the same bounds up to constants. Since the response values lie in \([0,1]\),
\[
 \Delta_n=\sup_{f\in L^2(I)}
          \|\widehat B_n(f)-\widehat B(f)\|_2\longrightarrow0
 \quad\text{almost surely}.
\]

Let \(L_0u=\mathbf 1_J M_\tau(\mathbf 1_A u)\) on \(L^2(I)\), where the output is
extended by zero off \(A\). The kernel bound gives
\(\|M_\tau(Cf-s^\star)\|_\infty\le c\|f-s^\star\|_2\) for a finite constant \(c\).
Thus, for \(f\) sufficiently near \(s^\star\) in \(L^2\), the output clip
of the continuum response uses the same regions \(Z,J,F\) as at \(s^\star\).
Furthermore
\(\|Cf-f\|_1\le\int_{\{|f-s^\star|\ge1\}}|f-s^\star|
 \le\|f-s^\star\|_2^2\), because \(s^\star\) takes values in \([0,1]\).
Boundedness of \(M_\tau:L^1(A)\to L^\infty(A)\) consequently gives
\[
 \|\widehat B(f)-s^\star-L_0(f-s^\star)\|_2
 \le c'\|f-s^\star\|_2^2
\]
for a finite constant \(c'\). The operator \(L=\mathrm{Id}-L_0\) is invertible: its equation
first fixes the coordinates outside \(J\), then solves with \(\mathrm{Id}-M_{\tau,JJ}\)
on \(J\).

Consider \(H_n(f)=L^{-1}(\widehat B_n(f)-L_0 f)\). This is a continuous
compact map on a closed ball about \(s^\star\), since \(\widehat B_n\) and
\(L_0\) are compact there. If \(\|f-s^\star\|_2\le r\) for sufficiently
small \(r\), then
\[
 \|H_n(f)-s^\star\|_2\le \|L^{-1}\|(\Delta_n+c'r^2).
\]
For \(\Delta_n>0\) and sufficiently large \(n\), take
\(r=2\|L^{-1}\|\Delta_n\). The bound places \(H_n\) within this ball, so
Schauder's fixed-point theorem gives a fixed point there. If \(\Delta_n=0\),
\(s^\star\) itself is fixed by \(\widehat B_n\). The equation \(f=H_n(f)\)
is equivalent to \(f=\widehat B_n(f)\). The finite equilibria constructed
above have distance \(O(\Delta_n)\) from \(s^\star\).
\end{proof}

\subsubsection*{Corner equilibria}

\begin{proof}[Proof of Proposition~\ref{pers:finite-bound}]
Apply Lemma~\ref{samp:inflows} to \(\mathcal F=\{\mathbf1_A,\zeta_0/V_0\}\). Both receipt
errors are at most \(E\) with failure probability at most
\(14n e^{-\ell}=\alpha\).
For the zero strategic profile the affine index changes from its sampled
continuum value by at most \(\|\nu\|_\infty E/p_{\min}\). For the full strategic
profile it changes by at most \(E/p_{\min}\), as actual and contractual
inflows coincide. Condition~\eqref{pers:finite-condition} keeps both corners
fixed, so they are the sampled extrema. If \(A_n=\varnothing\), the zero
profile is the only feasible profile and both corners coincide. Otherwise,
the finite mechanical operator
satisfies \((M_{0,n}R_n\zeta_0)_i\le \zeta_0(x_i)-1+V_0E/p_{\min}\le \zeta_0(x_i)-1/2\)
on active banks. Hence the mechanical response is a contraction in the weighted
maximum norm with weights \(\zeta_0(x_i)\). Its unique equilibrium is full early
payment, since full strategic repayment is feasible. Intersecting the
receipt-error event with \(A_n\ne\varnothing\) and using
\(\Pr(A_n=\varnothing)=(1-|A|)^n\) gives the joint probability bound.
\end{proof}

\begin{proof}[Proof of Corollary~\ref{pers:corners}]
Apply Lemma~\ref{samp:inflows} to \(\mathbf1_A\). The error in each sampled nominal
receipt tends to zero uniformly almost surely. Under
\eqref{pers:strict-corners}, the sampled affine index therefore remains
weakly below zero at the zero profile and weakly above one at the full
profile. These are the extremal strategic equilibria. If \(\rho(M_{0,\tau})<1\),
Theorem~\ref{samp:convergence} with \(h=\nu=0\) gives eventual almost-sure
mechanical uniqueness. The full profile is a mechanical fixed point
because full strategic repayment is feasible.
\end{proof}

\subsubsection*{Bank-level and aggregate approximation}

\begin{proof}[Proof of Theorem~\ref{samp:convergence}]
Put \(T=M_\tau\). Boundedness of \(P_\tau\), \(\bar p\) and \((1+\nu)/\bar p\) makes \(T:L^2(A)\to L^\infty(A)\) bounded. The spectral-radius assumption implies that \(\sum_{k\ge0}T^k\mathbf1_A\) converges in \(L^2(A)\). The same series converges in \(L^\infty(A)\) after its first term, by applying \(T:L^2(A)\to L^\infty(A)\) to its partial sums. Its sum \(\zeta\) satisfies \(\zeta\ge1\) and \(T\zeta=\zeta-\mathbf1_A\le\beta \zeta\), with \(\beta=1-1/\|\zeta\|_\infty\). Positivity of \(T\) and nonexpansiveness of clipping imply that \(B_\tau\) is a contraction in the norm \(\operatorname*{ess\,sup}_{x\in A}|u(x)|/\zeta(x)\), so \(s_\tau\) is unique.

Apply Lemma~\ref{samp:inflows} to \(\mathbf1_A,s_\tau,\zeta\), extending \(\zeta\) by zero off \(A\). Almost surely, eventually,
\[
 |B_n(R_ns_\tau)-R_ns_\tau|\le C_Rr_n\mathbf1_{A_n},\qquad
 M_nR_n\zeta\le(\beta+C_Mr_n)R_n\zeta
 \quad\text{on } A_n .
\]
The first bound follows from \eqref{samp:sample-response}, using \(K_nR_n\mathbf1_A\) for the nominal inflow. The second follows from \(T\zeta\le\beta \zeta\) and \(\zeta\ge1\). On the same event the sample map is a contraction in \(\max_{i\in A_n}|u_i|/\zeta(x_i)\) once \(\beta+C_Mr_n<1\). The weighted norm of its fixed-point error is bounded by the weighted residual of \(R_ns_\tau\) divided by \(1-\beta-C_Mr_n\). Multiplying by \(\|\zeta\|_\infty\) gives \eqref{samp:rate}. Both profiles are zero
at all sampled inactive types, so \eqref{samp:profile-bound} and
Lemma~\ref{samp:step} give the \(L^2\) limit and rate.
\end{proof}

\begin{proof}[Proof of Corollary~\ref{samp:aggregate}]
Replacing \(s^{(n)}\) by \(R_ns_\tau\) changes the sample ratio by at most the maximum error in \eqref{samp:rate}. Hoeffding's inequality and Borel--Cantelli give almost-sure convergence of the remaining ratio since \(\int_I\bar p(x)\,\mathrm dx>0\). The mechanical best response sets \(h=\nu=0\). Its positive linear part \(M_{0,\tau}\) is dominated by \(M_\tau\), hence has spectral radius at most \(\rho(M_\tau)\) by the operator norm comparison of their powers on \(L^2(A)\). Theorem~\ref{samp:convergence} and the same empirical-average argument then apply to the mechanical equilibrium.
\end{proof}

Lemma~\ref{samp:inflows} implies, for each fixed
bounded profile $f$,
\begin{equation}
 \|\mathcal U_n(R_nf)-R_n\mathcal U_\tau(f)\|_\infty=O(r_n)\quad\text{almost surely}.
 \label{pers:residual}
\end{equation}
The same lemma bounds the induced maximum norm \(\|M_n\|_\infty\) by a deterministic constant eventually almost surely. Indeed, the active row sums are \((1+\nu_i)\bar z_i^{(n)}/\bar p_i\), and
nominal inflows are uniformly close to their bounded continuum counterparts.

\begin{proof}[Proof of Proposition~\ref{pers:bracket}]
For a fixed integer $t$, apply \eqref{pers:residual} to the first $t$ continuum
iterates from $\mathbf 0$ and $\mathbf 1_A$. The maximum-norm Lipschitz bound
$\|M_n\|_\infty$ and the Lipschitz property of clipping give
$\|B_n^t(R_nf)-R_nB_\tau^t(f)\|_\infty=O(r_n)$ for
$f\in\{\mathbf 0,\mathbf 1_A\}$, almost surely by induction. Taking the intersection of these probability-one
 events over integer depths gives a single event for all fixed $t$. By monotonicity, these two
sampled iterates bound every sampled equilibrium from below and above. The
continuum iterates converge monotonically almost everywhere to their respective
limits, hence also in $L^1$. Boundedness of $P_\tau$, $\bar p$ and $\nu$, together
with the positive lower bound on active $\bar p$, gives
$\|B_\tau(u)-B_\tau(v)\|_\infty\le C\|u-v\|_1$ for a finite constant $C$. Applying \(B_\tau\) once
more gives uniform convergence in essential supremum. Choose $t$ so that the
continuum bounds are within $\varepsilon/2$ of their limits. The corresponding
almost-everywhere inequalities hold at the sampled types with probability one.
\end{proof}

\begin{proof}[Proof of Theorem~\ref{pers:branch}]
If $J$ is null, \eqref{pers:residual} preserves the strict saturation signs at
$R_ns^\star$. The bound on $\|M_n\|_\infty$ preserves them on a deterministic
neighbourhood, where the clipped response is the constant $R_ns^\star$.

Otherwise let \(w_J=\sum_{k\ge0}M_{\tau,JJ}^k\mathbf 1_J\) on \(J\) and extend it by
zero elsewhere. The series converges in \(L^2(J)\) by the spectral-radius condition.
It satisfies \(w_J=1+M_{\tau,JJ}w_J\) almost everywhere on \(J\), with \(w_J\ge1\). Since the kernel
is bounded, \(M_{\tau,JJ}\) maps \(L^2(J)\) into \(L^\infty(J)\), so
\(W_J=\|w_J\|_\infty<\infty\). Since \(J\) has positive measure, \(J_n\) is nonempty
eventually almost surely. The inflow approximation applied to \(w_J\) gives
the following inequality for the principal submatrix \((M_n)_{J_nJ_n}\), eventually almost surely:
\[
  (M_n)_{J_nJ_n}(R_nw_J)_{J_n}
      \le (R_nw_J)_{J_n}-\tfrac12\mathbf 1
      \le \left(1-\frac{1}{2W_J}\right)(R_nw_J)_{J_n}.
\]
Thus the interior submatrix has operator norm at most \(1-1/(2W_J)\) in the
weighted maximum norm \(\max_{i\in J_n}|v_i|/w_J(x_i)\). Fix repayments on $Z_n$ and
$F_n$ to zero and one and solve the affine equation on $J_n$. The inverse of
\(\mathrm{Id}-(M_n)_{J_nJ_n}\) exists and has weighted norm at most \(2W_J\). By
\eqref{pers:residual}, the resulting profile differs from $R_ns^\star$ in maximum
norm by $O(r_n)$ almost surely. It remains interior on $J_n$ and retains the required affine
signs on $Z_n,F_n$ by \eqref{pers:margins}, \eqref{pers:residual} and the row-sum
bound. It is therefore an equilibrium. For a sufficiently small deterministic
radius, every profile in the corresponding neighbourhood has those same affine
signs, including an affine index strictly between zero and one on $J_n$.
Every equilibrium there solves the same invertible system.

In either case, if $s^\star=s^-$, the least sampled equilibrium is below the constructed
equilibrium by monotonicity and above \(R_ns^\star-\varepsilon\mathbf 1_{A_n}\) eventually almost
surely by Proposition~\ref{pers:bracket}. It therefore lies in the same
neighbourhood eventually almost surely. The greatest case follows by
reversing the inequalities. Equation~\eqref{samp:profile-bound} and
Lemma~\ref{samp:step} give the $L^2$ convergence of the matched branch.
\end{proof}

\subsubsection*{Receipt dispersion and realised kernel norms}
\phantomsection
\label{samp:dispersion}

When \(\gamma\) stays fixed, the maximum over \(i\) of the absolute difference between the conditional variance in \eqref{samp:nominal-moments} and \((K_\tau\bar p)(x_i)/(\gamma+1)\) tends to zero in probability. The mean-weight part of Lemma~\ref{samp:inflows} applies with \(f=\bar p\) after rescaling, and the \(w_{ki}^2\) terms vanish because the maximum weight is of order \(1/n\). At types with positive limiting contractual claims, sampled claims retain positive conditional variance. Section~\ref{ex:fixed-gamma} shows how this dispersion can affect aggregate repayment.

Equilibrium consistency can coexist with a diverging squared \(L^2\) norm of the realised liability kernel. Let \(P_n\) be the step kernel with values \(n\Pi_{ki}\) on the sampled debtor--creditor cells of area \(1/n^2\). Then \(\|P_n\|_2^2=\sum_{k\in A_n}\sum_i\Pi_{ki}^2\). By \eqref{samp:dirichlet-moments} and \eqref{samp:mean-weights}, its conditional mean is \(|A_n|/(\gamma_n+1)+O(1)\) on the high-probability normalisation event. Its conditional variance is at most its mean, because the row concentrations are conditionally independent and lie in \([0,1]\). If \(\gamma_n/\log n\to\infty\) and \(\gamma_n=o(n)\), Chebyshev's inequality and \(|A_n|/n\to |A|\) give \(\|P_n\|_2^2/[n|A|/(\gamma_n+1)]\to1\) in probability. Creditors' receipts can therefore concentrate even when the debtor rows' squared share sums decay more slowly than the dense-row benchmark \(1/n\).

\subsection{The fixed-concentration limit}

\begin{proof}[Proof of Theorem~\ref{ex:fixed-limit}]
Let \(U_n,V_n,D_n\) be the sets of sampled banks in the three blocks and
put \(m=|U_n|\), \(\ell=|V_n|\). Hoeffding's inequality and Borel--Cantelli
give convergence of all block shares to \(1/3\) almost surely, including
for fresh draws at each \(n\). Eventually all three blocks are represented
and \(\ell\geq2\), so there are no fallback rows and the finite support is
acyclic. Fix \(j\in V_n\). Its early receipts are
\(Z_{j,n}=\bar z_j^{(n)}=\sum_{i\in U_n}\Pi_{ij}\), since upstream banks
pay one and owe one. Conditional on the types \(\boldsymbol{x}\), the
shares \(\Pi_{ij}\) for this \(j\) are independent across \(i\in U_n\), with
law \(\operatorname{Beta}(\gamma/\ell,\gamma-\gamma/\ell)\). Put
\(a=\gamma/\ell\) and write \(\mathrm{B}\) for the beta function. Then
\[
 \ell\,\mathbb E[1-e^{-t\Pi_{ij}}\mid\boldsymbol{x}]=
 \frac{\ell}{\mathrm{B}(a,\gamma-a)}
 \int_0^1 (1-e^{-tx})x^{a-1}(1-x)^{\gamma-a-1}\,dx .
\]
The prefactor tends to \(\gamma\) and the integrand is bounded by
\(t(1-x)^{\gamma/2-1}\) for \(\ell\geq2\). Dominated convergence, followed by
raising the single-row Laplace transform to the power \(m\), gives
\eqref{ex:fixed-laplace} along any count sequence with \(\ell\to\infty\)
and \(m/\ell\to1\).
The Poisson sum exists because \(\int x\,\mu_\gamma(dx)=1\), and its transform
follows by truncating jumps away from zero and taking the limit.

The conditional mean of \(\mathrm{RR}_n\) is
\((m+\ell\,\mathbb E[\min\{Z_{j,n},1\}\mid\boldsymbol{x}])/(m+\ell)\), which
converges to \(1-D_\gamma\) by the continuity theorem for Laplace transforms
and bounded continuity of \(z\mapsto\min\{z,1\}\) on \([0,\infty)\).
Changing one upstream row moves at most two units
of total downstream repayment, since the row shares sum to one and the cap is
1-Lipschitz. The conditional probability of a deviation of size \(\varepsilon\)
from the mean is thus at most \(2\exp[-\varepsilon^2(m+\ell)^2/(2m)]\).
This is summable when each block count lies within \(n/12\) of \(n/3\), as are
the probabilities of the complementary count event. Borel--Cantelli gives
the almost-sure limit in \eqref{ex:fixed-gap}.

The \(k\)-th cumulant of \(Z_\gamma\) is \(\gamma\mathrm{B}(k,\gamma)\).
In particular, the mean is one and the variance is
\(\sigma_\gamma^2=(\gamma+1)^{-1}>0\), so \(\Pr(Z_\gamma<1)>0\).
For \((Z_\gamma-1)/\sigma_\gamma\), Taylor expansion bounds the difference
between its log characteristic function and \(-t^2/2\) by
\(\lvert t\rvert^3\gamma\mathrm{B}(3,\gamma)/(6\sigma_\gamma^3)\to0\).
The normal limit and unit variance give convergence of the expected negative
part by uniform integrability, proving \eqref{ex:fixed-large}.
\end{proof}

\begin{proof}[Proof of Proposition~\ref{ex:fixed-order}]
Write \(\mu_\gamma(dx)=\gamma x^{-1}(1-x)^{\gamma-1}\mathbf1_{(0,1)}(x)\,dx\)
for the L\'evy measure. Its size-weighted measure \(x\mu_\gamma(dx)\) is the
probability law of \(B_\gamma\sim\operatorname{Beta}(1,\gamma)\). For the
nondecreasing L\'evy process \(X^{(\gamma)}\) whose time-one law is \(Z_\gamma\),
the generator for a Lipschitz payoff \(f\) at \(w\ge0\) is
\[
 \mathcal L_\gamma f(w)
 =\int_0^1\big(f(w+x)-f(w)\big)\mu_\gamma(dx)
 =\mathbb E\!\left[\frac{f(w+B_\gamma)-f(w)}{B_\gamma}\right].
\]
To verify the formula, first restrict jumps to \(x>\epsilon\) and differentiate
the compound Poisson expectation at time zero. Restoring the smaller jumps
changes the difference quotient in time by at most
\(\operatorname{Lip}(f)\int_0^\epsilon x\mu_\gamma(dx)\), uniformly in \(w\).
This tends to zero as \(\epsilon\downarrow0\).

For convex \(f\) the quotient in the expectation increases weakly with the
jump size. If \(\gamma_1<\gamma_2\), the tail formula
\(\Pr(B_\gamma>b)=(1-b)^\gamma\) for \(b\in[0,1]\) therefore yields
\(\mathcal L_{\gamma_1}f\ge\mathcal L_{\gamma_2}f\). Interpolate between the
time-one laws using independent processes and put
\(W_t=X_t^{(\gamma_1)}+X_{1-t}^{(\gamma_2)}\). Differentiating the two
convolution semigroups gives
\[
 \frac{d}{dt}\mathbb E f(W_t)
   =\mathbb E\big[(\mathcal L_{\gamma_1}-\mathcal L_{\gamma_2})f(W_t)\big].
\]
To see the derivative, for a small \(d>0\) condition both expectations on
\(X_t^{(\gamma_1)}+X_{1-t-d}^{(\gamma_2)}\). The remaining increments have time
length \(d\) in the two respective processes. The uniform time-zero generator
formula and stochastic continuity give the displayed derivative.
For \(f(w)=(1-w)_+\), the generator difference is strictly positive at
\(0<w<1\). Indeed, the quotient is \(-1\) for jumps up to \(1-w\) and is strictly
increasing thereafter, where the beta tails are strictly ordered. Each \(W_t\)
has mean one, positive variance and no atom at zero because its L\'evy measure
has infinite total mass. Thus \(\Pr(0<W_t<1)>0\), proving strict inequality of
the expectations at the endpoints.

Finally the log Laplace transform of \(Z_\gamma\) is
\(\mathbb E[(\exp(-uB_\gamma)-1)/B_\gamma]\) for \(u\ge0\).
As \(\gamma\downarrow0\), \(B_\gamma\) converges to one and the expression
converges to \(\exp(-u)-1\), the log Laplace transform of
\(\operatorname{Poisson}(1)\).
The bounded continuous payoff \((1-z)_+\) then gives the gap limit, since it
equals one at zero and vanishes at every positive integer.
\end{proof}

\begin{proof}[Derivation of \eqref{ex:fixed-unit}]
At \(\gamma=1\), the change of variable \(x=e^{-u}\) sends \(dx/x\) to unit
Lebesgue intensity. The first Poisson arrival and independence of the remaining
arrivals give \(Z_1\overset{d}{=}U_0(1+Z_1')\), with \(U_0\) uniform on \((0,1)\)
and \(Z_1'\) an independent copy of \(Z_1\). Integrating over \(U_0\) gives
\(D_1=\mathbb E[(1+Z_1)^{-1}]/4\). For \(L_1(t)=\mathbb E[e^{-tZ_1}]\),
\eqref{ex:fixed-laplace} implies \((tL_1(t))'=e^{-t}L_1(t)\), so
\[
 \mathbb E[(1+Z_1)^{-1}]=\int_0^\infty e^{-t}L_1(t)\,dt
 =\lim_{t\to\infty}tL_1(t).
\]
Integration by parts gives
\[
 \int_0^t\frac{1-e^{-u}}{u}\,du
 =(1-e^{-t})\log t-\int_0^t e^{-u}\log u\,du
 =\log t+\gamma_{\mathrm E}+o(1).
\]
The last equality uses \(\Gamma'(1)=-\gamma_{\mathrm E}\).
By \eqref{ex:fixed-laplace} the displayed limit is \(e^{-\gamma_{\mathrm E}}\).
\end{proof}

\clearpage
\suppressfloats[t]
\section{Sampling error and movement of the continuum target}
\label{ex:targets}

Use the four-block kernel from Section~\ref{ex:sharpening} and set \(e=9/20\), \(h=1/4\) and \(\nu=3/20\) on the debtor blocks. The active banks' common contractual claim \(r_\tau\) from
\eqref{ex:active-share} takes three values \((r_1,r_{16/25},r_{2/5})=(16/25,5/8,7/10)\). The continuum feedback radius \((23/20)r_\tau\) is less than one and the unique equilibrium is interior and symmetric. Its aggregate repayment rate is
\[
\mathrm{RR}_\tau=\frac{1/5-(3/20)r_\tau}{1-(23/20)r_\tau}
\quad\text{with}\quad
(\mathrm{RR}_1,\mathrm{RR}_{16/25},\mathrm{RR}_{2/5})=\left(\frac{13}{33},\frac{17}{45},\frac{19}{39}\right).
\]
The mechanical population benchmark is uniquely full at all three targets, since \(e+r_\tau>1\) and \(r_\tau<1\).

Let \(\mathrm{RR}_{n,\tau}\) be the aggregate equilibrium rate in a finite network sampled at \(\tau\) with a unique equilibrium. The signed difference from the original continuum prediction decomposes exactly as
\begin{equation}
 \mathrm{RR}_{n,\tau}-\mathrm{RR}_1
 = (\mathrm{RR}_{n,\tau}-\mathrm{RR}_\tau)+(\mathrm{RR}_\tau-\mathrm{RR}_1).
 \label{ex:repayment-decomposition}
\end{equation}
The sampling error \(\mathrm{RR}_{n,\tau}-\mathrm{RR}_\tau\) includes variation in the sampled population and realised exposures. It converges to zero almost surely under \(\gamma_n=4n\). The change in the continuum target \(\mathrm{RR}_\tau-\mathrm{RR}_1\) is zero at \(\tau=1\) and can have either sign under sharpening.

At both sharpened targets the predecessor link has zero weight in the population. At \(\tau=16/25\), the share owed to pure creditors rises from \(0.36\) to \(0.375\), reducing active banks' contractual claims and lowering the equilibrium repayment rate by \(1.62\) percentage points. Sharpening to \(\tau=2/5\) reduces the share owed to pure creditors to \(0.30\), increasing the repayment rate by \(9.32\) points relative to \(\mathrm{RR}_1\). The equilibrium rate increases in \(r_\tau\) throughout the interior regime. In this example, pruning the same set of links yields opposite changes relative to the original kernel because the remaining claims are allocated differently. The sampler fixes each debtor's total obligations and allows creditor totals to move, so the comparison includes changes in contractual claims.

\begin{figure}[!htbp]
 \centering
 \includegraphics[width=0.98\linewidth]{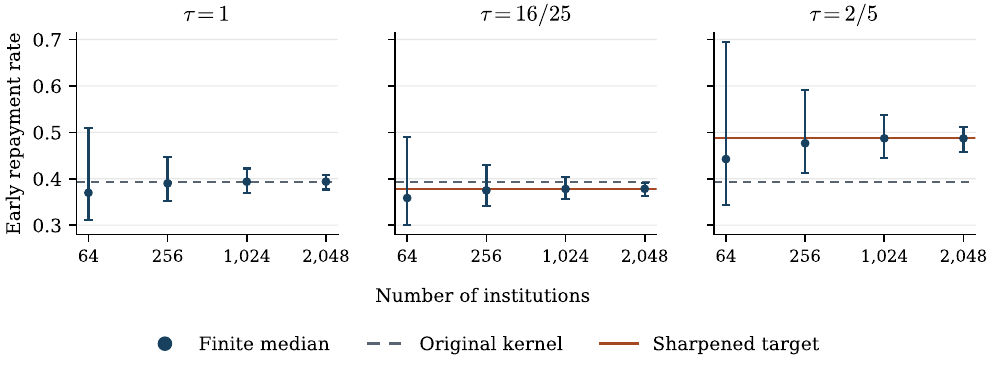}
 \caption{Finite repayment rates and continuum targets. Dots show the median and whiskers span the 10th to 90th percentiles across 48 draws at each size and sharpening parameter. The dashed line is the original continuum prediction \(\mathrm{RR}_1\). The solid line in each sharpened case is the corresponding target \(\mathrm{RR}_\tau\).}
 \label{ex:target-figure}
\end{figure}

Figure~\ref{ex:target-figure} reports 48 independent synthetic replications per size. Within each replication, all three values of \(\tau\) use the same sampled types and conditionally independent Dirichlet allocations, with \(\gamma_n=4n\). Each of the 576 finite games passed the contraction certificate \(\|M_n^k\|_\infty<1\) for some positive integer \(k\). Extremal iteration bracketed each equilibrium to within \(10^{-10}\) in maximum norm. Table~\ref{ex:sampling-table} separates the two terms at the largest sample size. The finite rates cluster around their own targets as size increases. In particular, the gap from the original prediction under stronger sharpening remains large even when the sampling error is small.

\begin{table}[!htbp]
 \centering
 \caption{Signed repayment differences at \(n=2{,}048\) in percentage points}
 \label{ex:sampling-table}
 \begin{tabular}{lrrrr}
 \toprule
 Sharpening \(\tau\) & \shortstack[r]{Change\\in target} & \shortstack[r]{Median\\sampling error} & \shortstack[r]{Median\\total difference} & \shortstack[r]{Median absolute\\sampling error} \\
 \midrule
\(1\) & \(0.00\) & \(-0.01\) & \(-0.01\) & \(0.88\) \\
\(16/25\) & \(-1.62\) & \(+0.06\) & \(-1.55\) & \(0.80\) \\
\(2/5\) & \(+9.32\) & \(-0.03\) & \(+9.30\) & \(1.56\) \\
 \bottomrule
 \end{tabular}
 \smallskip
 
 \parbox{0.94\linewidth}{\small The sampling and total differences are \(\mathrm{RR}_{n,\tau}-\mathrm{RR}_\tau\) and \(\mathrm{RR}_{n,\tau}-\mathrm{RR}_1\). Values are rounded independently. The seed is 831729, using NumPy's PCG64.}
\end{table}

\FloatBarrier
\section{Details for the four-block experiment}
\label{app:coexistence}

\paragraph{Calibration and computation.}
The calibrations with \(h>0\) can be generated using a unit exponential
normalised funding shock, \(\delta/\chi=\exp(-1)\), baseline funding scale
\(\omega=h\) and response coefficient \(\psi=\nu\). This gives \(c=1\).
The experiment in Section~\ref{ex:coexistence-simulation} samples 48 independent
networks per size and concentration, with uniform types and the
Sparsemax--Dirichlet allocation \eqref{samp:sampler}. Its first three
concentration regimes, \(4n,\sqrt n,(\log n)^2\), satisfy
\(\gamma_n/\log n\to\infty\). The fourth regime, \(2\log n\), has
\(\gamma_n\to\infty\), as required for \(L^2\) persistence under
nondegeneracy. The fifth holds concentration fixed at 12. We test the
corner conditions at their unclipped indices. When a corner fails,
monotone iteration starts at the corresponding endpoint, terminates
once the maximum update is below \(10^{-11}\), and verifies the
clipped residual.

To check the intermediate branch, we solve \((\mathrm{Id}-M_n)s=g_n\) on the active
debtor coordinates in the first eight draws in every size and concentration
cell. Here \(g_n\) and \(M_n\) are the intercept and linear part of the
sampled affine index before clipping. A candidate counts only when its active
coordinates lie in \((10^{-8},1-10^{-8})\) and the maximum absolute clipped
residual is at most \(10^{-9}\). At \(n=4{,}096\),
the median maximum deviation from \(103/187\) was \(0.0014\)
with \(\gamma_n=4n\) and \(0.0551\) with fixed \(12\).
In a typical draw, the projected rows offer about 2,050 potential
counterparties. Median effective counterparty counts at this size are
1,159.7 with \(4n\) and 14.1 with fixed \(12\).

\paragraph{Uniqueness of finite mechanical clearing.}
Let \(s_{\Phi,i}^\pm\) be the greatest and least mechanical repayment
fractions. In monetary units write \(p_{\Phi,i}^\pm=\bar p_i s_{\Phi,i}^\pm\)
and set \(d_i=p_{\Phi,i}^+-p_{\Phi,i}^-\). Let \(D_0=\{i:d_i>0\}\).
The mechanical cap is 1-Lipschitz in cash units, so
\(d_i\le\sum_j\Pi_{ji}d_j\). Summing over \(D_0\) forces every debtor in
\(D_0\) to allocate its entire row to \(D_0\). Every bank there has
\(p_{\Phi,i}^-<\bar p_i\) and strictly positive opening cash, so
\(p_{\Phi,i}^-=e_i+\sum_j\Pi_{ji}p_{\Phi,j}^-\).
Positive opening cash and nonnegative outside receipts imply payments
strictly greater than internal receipts when summed over \(D_0\).
Row closure makes them equal, a contradiction. Thus \(D_0\) is empty. We test full mechanical clearing by its cap condition
and use monotone iteration if that condition fails.

\paragraph{Failure of the exact corners under fixed concentration.}
Fix \(\gamma>0\) in the coexistence calibration. When the sampled block
masses are close to \(1/4\), the strict Sparsemax threshold eliminates the
weaker active link, and the total mean share to the successor block tends
to \(9/10\). Conditional on the types, each successor bank's contractual
claims are a sum of \(m\) independent
\(\operatorname{Beta}(\gamma r_n^{\mathrm{succ}}/\ell,
\gamma(1-r_n^{\mathrm{succ}}/\ell))\)
allocations, where \(m\) and \(\ell\) are the sizes of the predecessor and
successor blocks and \(r_n^{\mathrm{succ}}\) is the common projected share
allocated to the successor block by a predecessor row. Since
\(r_n^{\mathrm{succ}}\to9/10\) and \(m/\ell\to1\), the
Laplace-transform argument of Theorem~\ref{ex:fixed-limit} gives a limiting claim \(Z\) with
intensity \(\tfrac9{10}\gamma x^{-1}(1-x)^{\gamma-1}\,dx\)
on \(0<x<1\).

For any \(a>0\), \(\mathbb E[(a-Z)_+]>0\), since there is positive
probability of no jumps above a small cutoff. Independently, the sum below
the cutoff is less than \(a\) with positive probability by Markov's inequality
when the cutoff makes its mean less than \(a/2\). Changing one predecessor row
changes the block average of \((a-\bar z_i^{(n)})_+\) by at most \(2/\ell\).
The Efron--Stein inequality thus bounds its conditional variance by
\(2m/\ell^2\). Taking \(a=52/115\) shows that the block average converges
in probability to a positive number. With probability tending to one,
at least one active bank therefore has claims below \(52/115\), violating
the freeze condition \(\bar z_i^{(n)}\ge52/115\) and the full-payment condition
\(\bar z_i^{(n)}\ge12/25\) at that bank.

\paragraph{A partial freeze.}
Add a fifth block and give each block mass \(1/5\), preserving the original
outgoing block shares. The new block sends its debt only to the pure creditors
and has \(\bar p=1\), \(e=2\) and \(h=\nu=0\), which is consistent
with any \(\omega>0\) and \(\delta\geq\chi\). The three original debtors'
responses are unaffected. At \(\tau=7/20\) the least equilibrium has zero
early payment in the original debtor blocks and full payment in the new
block, for an aggregate early repayment rate of \(1/4\).

\FloatBarrier
\section{Fixed-concentration calculations}
\label{app:fixed-gamma}

In the three-block economy of Section~\ref{ex:fixed-gamma},
Theorem~\ref{ex:fixed-limit} gives the limiting shortfall
\(D_\gamma=\tfrac12\mathbb E[(1-Z_\gamma)_+]\), where \(Z_\gamma\) is the sum of
the points of a Poisson random measure on \((0,1)\) with intensity
\(\mu_\gamma(dx)=\gamma x^{-1}(1-x)^{\gamma-1}\,dx\). The receipt variable
\(Z_\gamma\) has mean one and variance \(\sigma_\gamma^2=1/(\gamma+1)\).

Let \(Y=Z_\gamma-1\). The fourth central moment satisfies
\[
 \mathbb E[Y^4]=3\sigma_\gamma^4+\gamma \mathrm{B}(4,\gamma)
 \leq 3\sigma_\gamma^4+\sigma_\gamma^2,
\]
where \(\mathrm{B}\) is the beta function. Since \(\mathbb E[Y]=0\),
\(D_\gamma=\mathbb E|Y|/4\). Hölder's inequality gives
\(\mathbb E|Y|\geq \sigma_\gamma^3/\sqrt{\mathbb E[Y^4]}\), yielding
\[
 D_\gamma\geq\frac{1}{4\sqrt{(\gamma+1)(\gamma+4)}}.
\]
The bound decays as \(\gamma^{-1}\), while the gap decays as \(\gamma^{-1/2}\).

\begin{table}[htbp]
 \centering
 \caption{Limiting early-payment shortfall under fixed concentration}
 \label{ex:fixed-numerical}
 \begin{tabular}{rrrr}
 \toprule
 \(\gamma\) & Gap \(D_\gamma\) & Limit \(1-D_\gamma\) & Moment lower bound \\
 \midrule
 1   & 0.14036 & 0.85964 & 0.07906 \\
 4   & 0.08860 & 0.91140 & 0.03953 \\
 16  & 0.04820 & 0.95180 & 0.01356 \\
 64  & 0.02471 & 0.97529 & 0.00376 \\
 100 & 0.01983 & 0.98017 & 0.00244 \\
 \bottomrule
 \end{tabular}

 \smallskip
 \parbox{0.90\linewidth}{\small Numerical entries use a discrete Fourier transform
 of the compound-Poisson law obtained by rounding each jump to adjacent grid
 points with its conditional mean preserved. The mesh is \(2^{-16}\).
 For a grid width \(d\) and \(\gamma\geq1\), the gap error from rounding is at most
 \(\tfrac{d}{2}\sqrt{\gamma(1+\log(1/d)/4)}\), by integrating the rounding variance
 against \(\mu_\gamma\). At grid period \(16\), the wrap-around error is at most
 \(\tfrac12\exp(15-16\log 16)\), since each rounded jump is at most one and the
 sum has mean one. Together these bounds are below \(0.00015\) for the values
 shown. The discrete intensities are integrated numerically using
 Gauss--Legendre quadrature. The replication code checks agreement across
 two grid resolutions and quadrature orders.}
\end{table}

\subsection{Confidence-sensitive reserves}

When wholesale counterparties in \(U\) pay in full, their payments to the
dealers in \(V\) equal the dealers' contractual claims, so there are no
missed receipts on that link. Set \(h_U=1-q\) so that the counterparties pay
\(q\in(0,1)\), and keep \(e_U=e_V=1\). Set \(a=1-h_V\in[0,1)\) and
\(0<\nu_V=\nu<q/(1-q)\), so \(b=q-\nu(1-q)>0\). A dealer owed
\(\bar z\) receives \(q\bar z\) and raises desired reserves by \(\nu(1-q)\bar z\). It pays
\(\min\{1,a+b\bar z\}\). The bounded-differences argument for aggregate repayment
applies to this capped linear function. The limiting finite repayment and the
continuum repayment are, respectively,
\[
 \tfrac12\bigl(q+\mathbb E[\min\{1,a+bZ_\gamma\}]\bigr),
 \qquad
 \tfrac12\bigl(q+\min\{1,a+b\}\bigr).
\]
By concavity, the sampled limit is strictly below the continuum rate:
\(Z_\gamma\) has positive probability on both sides of the cap threshold
\((1-a)/b>0\).
For its upper tail, arbitrarily many Poisson jumps can occur in a fixed
interval bounded away from zero. For its lower tail, exclude jumps above a
small cutoff and apply Markov's inequality to the independent sum of smaller
jumps, whose expectation tends to zero with the cutoff.

Fix \(a=1-q\) and vary \(\nu\) within the stated range at fixed \(q\).
Continuum dealers repay \(1-\nu(1-q)\), and the aggregate sampling gap is
\(\tfrac12\mathbb E[(bZ_\gamma-q)_+]\). Raising \(\nu\) lowers \(b\), reducing both
the continuum rate and the sampling gap. Choosing \(a=1-b\) puts continuum
dealers exactly at their cap and gives gap \(bD_\gamma\). The gap remains
positive for \(1-b<a<1\), when their continuum resources strictly exceed
the cap.

\FloatBarrier
\section{Smooth-kernel diagnostics}
\label{app:cosine-diagnostics}

Let \(A=[0,0.8]\) and \(P(x,y)=1+0.5\cos(2\pi(x-y))\) on active rows,
with \(P=0\) off \(A\). On \(A\), set \(\bar p(x)=1+x\) and \(e(x)=0.3\).
The normalised funding shock is generalised Pareto with shape \(1/2\)
and unit scale. With \(\chi=1.2\), \(\delta=1\) and \(\omega=\psi=0.5\),
the reserve and receipt-sensitivity coefficients are
\(h=\nu=\sqrt{1.2}-1\). We use the canonical sampler at \(\tau=1\) and
use the realised contractual claims \(\bar z_i^{(n)}\) in the sampled
best response \eqref{samp:sample-response}.

The cosine addition formula reduces continuum inflows to three moments
of the monetary payment profile. In the norm
\(\|f\|_{1,\bar p}=\int_A\bar p(x)|f(x)|\,\mathrm dx\), the continuum feedback
operator has norm bounded by
\((1+\nu)(0.8+\sin(0.8\pi)/(2\pi))<1\), so the continuum equilibrium is
unique. The 200,000-point midpoint rule gives
\(\mathrm{RR}_1=0.705867\), and doubling the grid changes the
evaluated repayment profile by less than \(10^{-10}\) on a uniform
check grid.

Table~\ref{app:cosine-table} separates two experiments with 12
independent draws per specification. At \(n=3{,}200\), Panel A reports the
mean signed difference \(\mathrm{RR}_{n,1}-\mathrm{RR}_1\) and the standard
deviation of the sampled rates \(\mathrm{RR}_{n,1}\). Panel B reports the mean
of \(\max_{i\in A_n}|s_i^{(n)}-s_1(x_i)|\) at increasing sizes with
\(\gamma_n=n\), illustrating Theorem~\ref{samp:convergence}. Weighted contraction certifies uniqueness in all 96
draws. Iteration from the two lattice endpoints brackets each numerical
repayment profile to within \(10^{-9}\) in maximum norm. The code uses
PCG64 with seed 548902 and records the draws and summary statistics.

\begin{table}[htbp]
\centering
\caption{Aggregate repayment and bank-level errors under a smooth kernel}
\label{app:cosine-table}
\begin{minipage}[t]{0.49\linewidth}
\centering
\textit{Panel A. Fixed size \(n=3{,}200\)}\par\smallskip
\begin{tabular}{rrr}
\toprule
\(\gamma\) & Mean shift & Rate SD \\
\midrule
\(5\)         & \(-0.124\) & \(0.023\) \\
\(20\)        & \(-0.049\) & \(0.030\) \\
\(5\sqrt n\)  & \(-0.015\) & \(0.029\) \\
\bottomrule
\end{tabular}
\end{minipage}\hfill
\begin{minipage}[t]{0.49\linewidth}
\centering
\textit{Panel B. Growing size, \(\gamma_n=n\)}\par\smallskip
\begin{tabular}{rr}
\toprule
\(n\) & Mean maximum error \\
\midrule
\(200\)   & \(0.225\) \\
\(400\)   & \(0.170\) \\
\(800\)   & \(0.142\) \\
\(1{,}600\) & \(0.119\) \\
\(3{,}200\) & \(0.081\) \\
\bottomrule
\end{tabular}
\end{minipage}
\end{table}

\FloatBarrier
\section{Synthetic core--periphery stress test}
\label{app:core-periphery}

\paragraph{Calibration.}
In the stress test of Section~\ref{ex:core-periphery}, let \(C,R,D\)
denote the core, periphery and cash creditors.
Their masses are \(1/10,7/10,1/5\). The marks and the constant affinity
\(Q(x,y)\) on each debtor--creditor rectangle are given in
Table~\ref{app:core-periphery-marks}. In \(Q(x,y)=W(\phi(x),\varphi(y))\),
\(\phi\) and \(\varphi\) return the sender and receiver group labels,
respectively, and \(W\) is the table of affinity scores. All quantities are
in normalised units. The parameters and counterparty counts are illustrative.

\begin{table}[!htbp]
\centering
\caption{Core--periphery marks and directed affinity scores}
\label{app:core-periphery-marks}
\small
\begin{tabular}{@{}lrrrrrrr@{}}
\toprule
Debtor group & \(\bar p\) & \(e\) & \(h\) & \(\nu\) & \(Q_{\cdot C}\) & \(Q_{\cdot R}\) & \(Q_{\cdot D}\) \\
\midrule
Core       & 4 & 3.95 & 2.05 & 1.55 & 3.0 & 0.6 & 1.4 \\
Periphery  & 1 & 1.10 & 0.05 & 0    & 1.1 & 0.9 & 1.3 \\
\bottomrule
\end{tabular}
\par\smallskip
\parbox{0.96\linewidth}{\small The creditor group \(D\) has \(\bar p=0\) and a zero affinity row. Each active row integrates to one, so the starting relative-liability kernel is \(P_1=Q\).}
\end{table}

\paragraph{Population benchmarks.}
The response is constant within each block, and the periphery always
pays one under the stated marks. Write \(a s_C+b\) for a core bank's
receipts when the core pays fraction \(s_C\). Here \(a\) and \(b\) are
the claims on the core and periphery, respectively, per core bank.
The core response is
\[
 \min\{1,\max\{0,(e_C-h_C+b-\nu_C a+(1+\nu_C)a s_C)/\bar p_C\}\}.
\]
At \(\tau=1\), \((a,b)=(6/5,77/100)\), giving a contraction and
\(s_C=81/94\). At \(\tau=1/5\), Sparsemax gives outgoing block shares
\((13/15,0,2/15)\) and \((3/20,7/20,1/2)\) for the core and periphery,
respectively. Then \((a,b)=(52/15,21/20)\). The zero core repayment profile
is a fixed point because the unclipped intercept is negative. Full
payment is a fixed point because the unclipped index at one exceeds one.
Thus \((0,1)\) and \((1,1)\) are the least and greatest repayments of the
two debtor groups. The core and periphery owe \(2/5\) and \(7/10\) in total,
respectively. The least aggregate repayment rate is \(7/11\).

The sharpened target claims per bank are \((271/60,7/20,121/60)\).
The maximum-entropy benchmark has the same claim and liability margins.
Set \(P^{\mathrm{ME}}(x,y)=\bar z_{1/5}(y)/(11/10)\) on active rows and
zero elsewhere. Normalising nominal liabilities by their total turns them
into a joint density with specified debtor and creditor marginals. Its
cross-entropy with the product of those marginals is fixed. Relative
entropy is minimised at that product, giving the stated benchmark. In the
resulting core response, \(a=271/165\) and \(b=1897/660\). Its intercept is
positive and its slope exceeds one, so the unique strategic equilibrium
is full. All three population benchmarks have unique full mechanical
clearing. With reserves suppressed, peripheral payments alone make full
core payment feasible in each of them.

\paragraph{Draws and sensitivity.}
Each network contains 60 core banks, 420 periphery banks, and 120 cash
creditors. We condition on these counts, mask the diagonal, and apply the
row normalisation, Sparsemax and Dirichlet steps of \eqref{samp:sampler}.
For each finite-concentration specification we generate 120 independent
allocations with NumPy's PCG64 and seed sequence \((274339,j,k)\), where
\(j\) indexes the scenarios in script order and \(k\) indexes draws, both
starting at zero. At \(\gamma=\infty\) the projected weights give a single
deterministic allocation. Monotone iteration from zero and full payment gives numerical extrema
with maximum clipped residual at most \(10^{-11}\). All 602
matrices satisfy the full mechanical clearing condition at every debtor.
Strictly positive opening liquidity ensures mechanical uniqueness by
the argument in Appendix~\ref{app:coexistence}.
Table~\ref{app:core-periphery-sensitivity} reports the sensitivity to
Sparsemax temperature and Dirichlet concentration.

\begin{table}[!htbp]
\centering
\caption{Sensitivity of the synthetic finite-network stress test}
\label{app:core-periphery-sensitivity}
\small
\setlength{\tabcolsep}{4pt}
\begin{tabular}{@{}rrrrrr@{}}
\toprule
\(\tau\) & \(\gamma\) & \shortstack[r]{Least rate\\5th / 50th / 95th} & \shortstack[r]{Greatest\\median} & \shortstack[r]{Effective\\count} & \shortstack[r]{Claims\\deviation} \\
\midrule
1    & \(\infty\) & \mbox{94.67/94.67/94.67} & 94.67 & 559.40 & 66.26 \\
1    & 16        & \mbox{92.07/94.10/95.90} & 94.10 & 17.78  & 67.58 \\
0.30 & 16        & \mbox{63.84/64.19/64.72} & 99.99 & 17.40  & 25.79 \\
0.20 & \(\infty\) & \mbox{63.64/63.64/63.64} & 100.00 & 327.42 & 0.59 \\
0.20 & 8         & \mbox{63.79/64.13/64.68} & 100.00 & 9.71   & 26.87 \\
0.20 & 16        & \mbox{63.64/63.72/63.93} & 100.00 & 17.18  & 19.86 \\
0.20 & 32        & \mbox{63.64/63.64/63.67} & 100.00 & 30.75  & 14.22 \\
\bottomrule
\end{tabular}
\par\smallskip
\parbox{0.96\linewidth}{\small Repayment rates and claim deviations are in percent. Effective count is the median across draws of the active-row mean of \(1/\sum_j\Pi_{ij}^2\). Claims deviation is the median of \(\sum_i|\bar z_i^{(n)}-\bar z_{1/5}(x_i)|/\sum_i\bar p_i\), with both sums over all banks and the same target in every row. At \((1/5,16)\), the medians of within-group row means are \(15.02\) for the core and \(17.49\) for the periphery.}
\end{table}

\end{document}